\documentclass[aps,prx,10pt,notitlepage,twocolumn,superscriptaddress,nolongbibliography]{revtex4-2}
\usepackage{amsmath}
\usepackage{amssymb}
\usepackage{bbold}
\usepackage{color}
\usepackage{tikz}
\usepackage{chngcntr}
\usepackage{pgfplots}
\pgfplotsset{compat=1.5}
\usepackage{graphicx}

\definecolor{dodgerblue}{HTML}{1E90FF} 
\usepackage[colorlinks=true,citecolor=dodgerblue,linkcolor=dodgerblue,urlcolor=dodgerblue]{hyperref}
\usepackage{blindtext}
\usepackage{physics}

\newcommand{\br}{{\bf r}}
\newcommand{\bk}{{\bf k}}
\newcommand{\bq}{{\bf q}}
 
\newcommand{\moire}{moir\'e}

\def\bk{{\bf k}}

\def\ba{{\bf a}}

\def\bq{{\bf q}}

\def\bR{{\bf R}}

\def\bG{{\bf G}}
\def\bd{{\bf d}}

\def\bA{{\bf A}}

\def\bd{{\bf d}}
\def\br{{\bf r}}
\def\bq{{\bf q}}

\def\bsigma{{\boldsymbol \sigma}}

\def\phdag{{\phantom \dagger}}

\begin{document}
\title{
Cell Natural Orbitals in Interacting Topological Bands
}
\author{Nishchhal Verma}
\email{nishchhal.verma@columbia.edu}
\affiliation{Department of Physics, Columbia University, New York, NY 10027, USA}

\author{Harshitra Mahalingam}
\affiliation{Department of Physics, Columbia University, New York, NY 10027, USA}

\author{Daniel Mu\~{n}oz-Segovia}
\affiliation{Department of Physics, Columbia University, New York, NY 10027, USA}

\author{Raquel Queiroz}
\email{raquel.queiroz@columbia.edu}
\affiliation{Department of Physics, Columbia University, New York, NY 10027, USA}
\affiliation{Center for Computational Quantum Physics, Flatiron Institute, New York, NY 10010, USA}

\begin{abstract}
Topological bands exhibit obstruction to exponentially localized and symmetric Wannier functions, challenging the standard paradigm of representing projected interactions in terms of local orbitals with finite range. 
To faithfully capture the form factors and quantum geometry of topological bands we introduce a singular-value decomposition of the band-projected density form factors, enabling a geometry-based truncation scheme of the Hilbert space, exposing an intrinsic hierarchy on band-projected interactions that is  determined by the underlying wavefunctions. This decomposition is most naturally described in terms of Cell Natural Orbitals (CNOs) \cite{PaperCNO1}, as the eigenstates of the unit-cell reduced one-particle density matrix, whose occupation
provide a measure of the minimal orbital complexity required to faithfully represent the band wavefunctions overlaps. 
The CNO decomposition identifies systematically the minimal number of local orbitals needed to reproduce short-ranged interactions while resolving the hierarchy of interaction strengths across CNO channels.
Applied to magic-angle twisted bilayer graphene in the chiral limit, we find that the dominant CNO is centered at the AA site, resembling the $f$-fermion of the heavy-fermion model \cite{song2022magic}.
The subdominant CNO channels carry progressively weaker interaction matrix elements, allowing them to be treated at the static mean-field level, while the dominant channel requires a dynamical self-energy.
The formalism illustrates how variations of charge density within the unit cell generate momentum dependence in the CNO envelope function and, consequently, dispersion in the single-particle spectral function.
More broadly, our results establish CNOs as a geometry-informed bridge between band topology and real-space correlations, providing a systematic framework for analyzing interactions and emergent phases in quantum materials.
\end{abstract}

\maketitle

\section{Introduction}
A recurring challenge in condensed matter physics is choosing the appropriate representation for an electronic system.
The momentum-space formalism of band theory is natural for describing topological phases, whereas the real-space language of local orbitals is better suited for describing interactions.
Interacting topological materials bring this tension into sharp focus.
Global properties of Bloch wavefunctions obstruct the construction of exponentially localized symmetric Wannier functions~\cite{BrouderPRL2007,Monaco2018,Po2017Indicators,Po2018Origin,SoluyanovPRB2011}. 
This complicates the traditional bridge between momentum-space band theory and local real-space interactions.
The recent proliferation {\moire} heterostructures~\cite{Andrei2021,CaoNature2018,CaoNature2018a, Lu2019_tbg, Yankowitz2019_tbg} makes this question urgent. How does one represent interactions when topology forbids exponentially localized Wannier functions?

In addition to band topology \cite{Hasan2010_TI}, quantum geometry is now recognized as an organizing principle for correlated materials \cite{Yu2025_npj, Verma2026_nrr}.
The quantum geometric tensor appears in various optical responses and sum rules \cite{provost1980riemannian, Torma2023_prl_commentary, verma2021, mao2023, Verma2025_pnas}.
Whether these results survive in the presence of interactions is less settled.
The quantum geometric tensor captures overlaps $\langle u_\bk | u_{\bk+\bq} \rangle$ (where $|u_\bk\rangle$ is the cell-periodic part of the Bloch wavefunction) at infinitesimal $\bq$, while projected interactions involve overlaps at an arbitrary momentum transfer.
Large-$\bq$ contributions probe the short-distance charge distribution within the unit cell and are not constrained by the quantum metric.
Therefore, whether the quantum metric in particular is the true driver of the effects attributed to it, or an artifact of approximations used to treat the large-$q$ aspect of interaction, is not settled.

The discovery of {\moire} materials has highlighted this
issue further ~\cite{Lopes2007,Shallcross2010,bistritzer2011moire,Lopes2012,
TBG1,Zou2018,song2019all,TBG2,Koshino2018,Kang2018,TBG3}.
The flat bands in these materials carry nontrivial quantum geometry and host strongly correlated phases~\cite{CaoNature2018,CaoNature2018a,
Lu2019_tbg},
making quantum geometry an essential ingredient in their phenomenology.
However, it is largely unclear which aspects of the wavefunction control the observed correlated phases.
Hartree-Fock calculations capture the full form-factor
structure~\cite{Xie2020,Zhang2020,Bultinck2020} but offer limited
physical insight into which parts matter.

\begin{figure*}
    \centering
    \includegraphics[width=1.95\columnwidth]{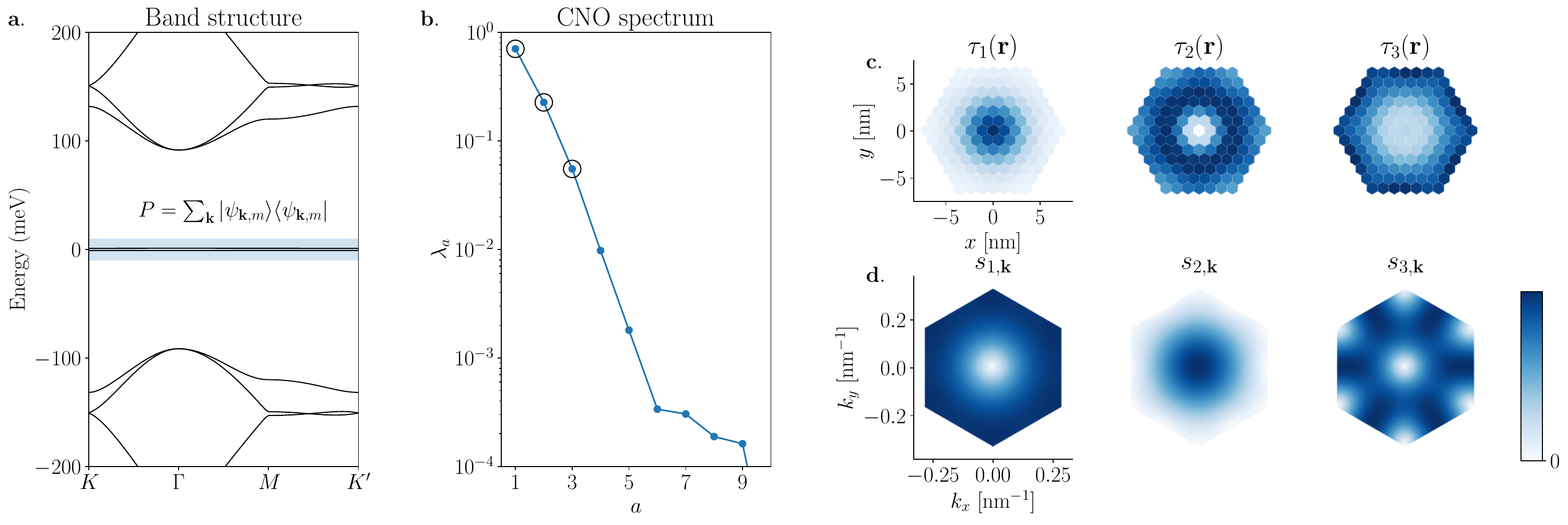}
    \caption{\textbf{Cell Natural Orbitals.}
    {\bf a}. Band structure of twisted bilayer graphene in the chiral limit with twist angle $\theta = 1.07^\circ$. There are two flat bands related by chiral symmetry. We select one of them to construct the low-energy projector $P$.
    {\bf b}. CNO spectrum obtained from the unit-cell reduced density matrix defined in Eq.~\eqref{eqrho_cell_BZ_average}. The spectrum decays exponentially, hinting that only a few orbitals are required to efficiently capture the charge density within the unit cell.
    {\bf c}. The first three CNOs with dominant occupation in real space. 
    The AA site is at the center of the unit cell while AB and BA sites form the corners.
    The dominant CNO is centered at the AA site in the {\moire} unit cell, reminiscent of the $f$ orbital in the heavy fermion model of TBG. The other CNOs are more delocalized.
    {\bf d}. Projection of the CNO onto the band wavefunction $s_{a,\bk}$, defined in Eq.~\eqref{eq:envelope_def}. There are zeros in the envelope function that are topologically protected. The first mode has a zero at the $\Gamma$ point.}
    \label{fig1}
\end{figure*}

Our current understanding of form factors in {\moire} materials is largely based on two complementary perspectives.
The first is the Landau-level analogy, which interprets flat bands as arising from spatially varying magnetic fields~\cite{
ledwith2020fractional,crepel2024topologically,Parhizkar2024,
liu2019pseudo, wang2021exact,ledwith2020fractional}.
This approach recasts the form factors in terms of Landau-level wavefunctions.
Even though the mapping is exact only in certain limits \cite{tarnopolsky2019origin, Shi2024_adiabatic} and is not universal across {\moire} materials, it provides useful intuition about fractional states that appear in certain materials.
The other, more universal approach, is representation in terms of a multi-orbital model, such as the topological heavy-fermion scheme \cite{song2022magic}. It decomposes the wavefunctions into localized $f$ and itinerant $c$
electrons~\cite{Kang2021H,Shi2022,Datta2023Heavy},
reproducing the flat-band form
factors~\cite{Calugaru2023}.
While highly successful in various {\moire} platforms, the identification of the heavy fermion orbital typically relies on an ansatz for the localized $f$ orbital. Ideally, the relevant local orbital should instead emerge directly from the underlying Bloch wavefunctions as an intrinsic feature of the band manifold.

What is currently lacking is a framework for generating local orbitals from low-energy bands while ranking their relative importance.
Such a framework would determine the minimal set of orbitals required to represent the form factors, outline
the constraints imposed by topology on interactions, and establish a decomposition of form factors enabling controlled truncations in orbital space.
The resulting local representation should be exact when all orbitals are retained and improvable upon truncation.

As introduced in Ref.~\cite{PaperCNO1}, the one-particle reduced density matrix restricted to a single unit cell provides such orbitals.
Its eigenfunctions, \textit{Cell Natural Orbitals} (CNOs), describe charge fluctuations within a given band or set of bands. 
CNOs are uniquely defined by the band structure and are well-localized by construction, even when topological obstructions are present.
Its eigenvalues (occupation numbers) quantify the complexity of the low-energy bands, determining the number of local orbitals needed to describe the charge density of the band(s).

Crucial to the interaction phenomenology, CNOs reveal a hierarchical structure in the density form factors. 
The leading CNO captures the dominant contribution to the local charge density, while subleading modes encode charge fluctuations enforced by the non-trivial quantum geometry of the band. This hierarchy not only enables a geometric truncation scheme for approximating projected interactions, but also highlights an intrinsic momentum scale associated with the spatial structure of the charge distribution within the unit cell. In particular, because the density is distributed among distinct CNOs, its envelope acquires a non-trivial momentum dependence that feeds into the single-particle spectral function.

Our framework offers conceptual insights beyond numerical utility.
The occupation spectrum also quantifies \textit{unit-cell entanglement} corresponding to the low-energy subspace, capturing correlations between the unit cell and the extended crystal.
This entanglement evolves characteristically across topological phase transitions, analogous to entanglement entropy in many-body systems \cite{Horodecki_x4,Amico2008}.
Bands requiring multiple CNOs with comparable occupations signal real-space entanglement between different unit-cell degrees of freedom.

The remainder of this paper is organized as follows. Section~\ref{sec1pRDM} uses the one-particle reduced density matrix, defines cell natural orbitals and their occupation numbers, and connects them to unit-cell entanglement. 
Section~\ref{secformfactors} derives the CNO decomposition of density form factors and identifies topologically protected zeros. 
Section~\ref{secCNObasis} uses CNOs to simplify projected interactions.
Section~\ref{sec:comparison} compares CNOs to existing methods and
Section~\ref{secconclusions} discusses broader implications for correlated topological materials.

\section{Cell Natural Orbitals}
\label{sec1pRDM}

We briefly collect the definitions and properties of cell natural
orbitals (CNOs) needed for the form-factor and interaction analysis
below. We refer the read to Ref.~\cite{PaperCNO1} for the full derivation.

Consider a periodic lattice with Bravais vectors $\bR$ and $N_\alpha$
internal degrees of freedom per unit cell (orbital, sublattice, spin,
layer), labeled $\alpha$. For a non-interacting state restricted to a
chosen set of bands with band projector $P$, define the
$N_\alpha\times N_\alpha$ matrix with elements 
\begin{equation}
L_{\alpha\beta}
=
\frac{1}{N_k}\sum_{\bk\in\mathrm{BZ}}
\sum_{n\in P}
u^{\vphantom{*}}_{n,\alpha}(\bk)\,u^*_{n,\beta}(\bk),
\label{eqrho_cell_BZ_average}
\end{equation}
where $N_k$ is the number of discrete Brillouin-zone momenta (equivalently, the number of unit cells), and
$u_{n,\alpha}(\bk)$ are the cell-periodic parts of the Bloch wavefunction in the periodic embedding, $u_{n,\alpha}(\bk+\bG)=u_{n,\alpha}(\bk)$. The periodic embedding can always be defined (for continuum models, it is subtle and its derivation is outlined in Appendix~\ref{app:periodic_continuum}). 
Throughout this paper we use $|u_{\bk}\rangle$ to denote Bloch states in the periodic embedding unless specified otherwise.
Eq.~\eqref{eqrho_cell_BZ_average} is the Brillouin-zone average
of the band projector $P$, restricted to a single unit cell.
It is Hermitian, positive semidefinite, gauge invariant and is independent of the embedding of orbitals (the equations for $L$ in embeddings other than the periodic one are modified accordingly).

Restricted correlation matrices and their natural occupations are standard in free-fermion block-density-matrix constructions \cite{Cheong2004}, closely related local occupation matrices are used in DFT+U \cite{Dudarev1998, Cococcioni2005}. Here the restricted region is the internal Hilbert space of one unit cell and the projector is a selected band manifold.
The restriction leads to fractional occupations even in a free-fermion description.

\begin{figure*}
    \centering
    \includegraphics[width=1.8\columnwidth]{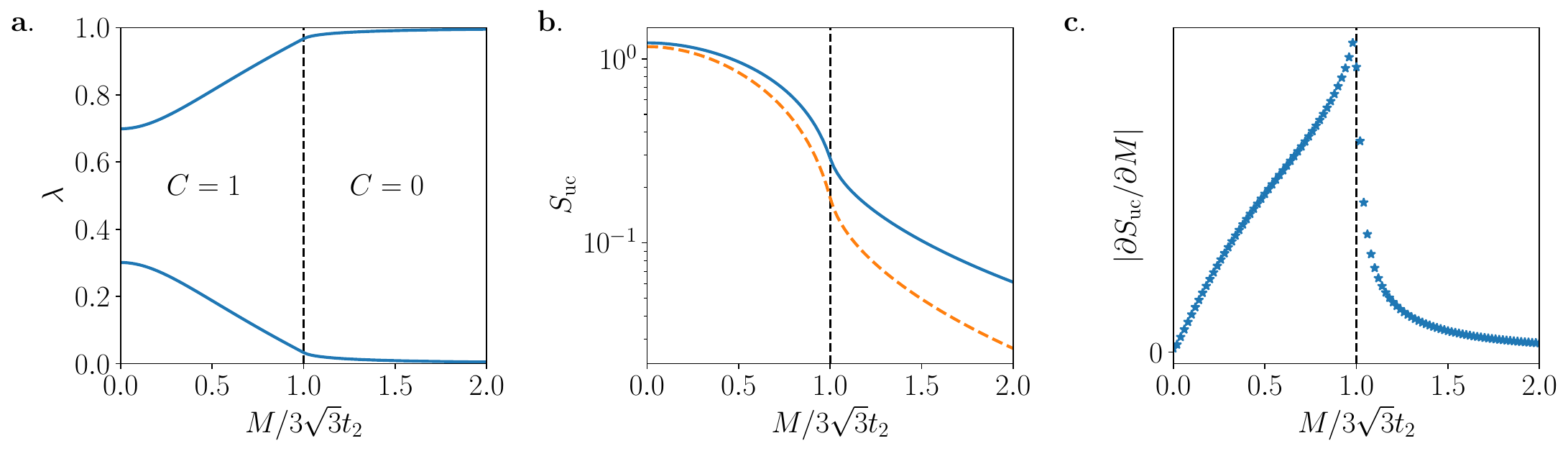}
    \caption{\textbf{Unit-cell entanglement across the Haldane model
    phase transition.}
    (a) CNO eigenvalues $\lambda_\pm$ vs staggered potential $M/t$
    at fixed flux $\phi=\pi/2$.
    (b) Unit-cell entanglement entropy $S_{\mathrm{cell}}$ showing
    finite entanglement in the topological phase
    ($M<3\sqrt{3}\,t_2\sin\phi$) and and entanglement that becomes exponentially small on approaching
    the atomic limit $M/t1 \rightarrow \infty$ in the trivial phase. $S_{\mathrm{cell}}$ is strictly positive for any finite $M$ and vanishes only in the atomic limit. The orange dashed line shows the lower-bound coming from average of quantum distance between Bloch states (defined in Eq.~\eqref{eq:Scell_bound_mt}).
    (c) The peak in $dS_{\mathrm{cell}}/dM$ marks the transition through
    a gapless Dirac point.}
    \label{fighaldane_model}
\end{figure*}

We pause here to note that technically, $L$ is the correlation matrix, which is related, but not identical, to the fermionic reduced density matrix $\rho_{\rm red}$. The latter is a tensor product over occupied/empty mode sectors $\rho_{\rm red} = \bigotimes_a \lambda_a |1_a \rangle \langle 1_a | + (1-\lambda_a) |0_a \rangle \langle 0_a |$, a relation first identified by Peschel \cite{Peschel2003,Peschel2009}. Since both $\rho_{\rm red}$ and $L$ are diagonalized by the same single-particle orbitals, we use them interchangeably in the discussion. The additional $(1-\lambda)$ contribution arises because the fermionic reduced density matrix acts in Fock space, where each mode has a two-dimensional local Hilbert space corresponding to occupied and empty states.

Diagonalizing the \emph{unit-cell reduced density matrix} (uc-RDM) yields the CNOs
$\{|\tau_a\rangle\}$ and \emph{natural occupations} $\{\lambda_a\}$,
\begin{equation}
L|\tau_a\rangle = \lambda_a|\tau_a\rangle,
\quad
\lambda_1\ge\lambda_2\ge\cdots\ge 0,
\label{eqCNO_eigenvalue}
\end{equation}
forming an orthonormal basis for unit-cell Hilbert space ordered by decreasing occupation.
The eigenvalues satisfy $\sum_a\lambda_a = \langle N_\mathrm{cell}\rangle$.
Unless specified, we will normalize the eigenvalues so that $0 \le \lambda_a \le 1$ with $\sum_a\lambda_a=1$.
CNOs are local by construction as they are linear combinations of the on-site orbitals and carry zero weight outside the unit cell, and are uniquely determined by $P$ without any gauge fixing.

The occupation numbers $\{\lambda_a\}$ quantify
\emph{unit-cell entanglement},
the degree to which the single-particle wavefunction cannot be
factored into independent unit-cell contributions.
Since $L$ is the single-particle correlation matrix of the free-fermion ground state restricted to the unit-cell degrees of freedom, the Peschel relation \cite{Peschel2003,Peschel2009} relates it to the entanglement Hamiltonian of the unit-cell bipartition, with entanglement
energies $\varepsilon_a = \log[(1-\lambda_a)/\lambda_a]$.
The von Neumann entropy of the unit-cell bipartition is
\begin{equation}
S_{\mathrm{cell}}
=
-\sum_a\Bigl[\lambda_a\ln\lambda_a
+(1-\lambda_a)\ln(1-\lambda_a)\Bigr].
\label{eqS_cell}
\end{equation}
Modes with $\lambda_a\in\{0,1\}$ are unentangled;
$\lambda_a = 1/2$ gives maximal single-mode contribution $\ln 2$.
The cell entanglement is symmetric under $\lambda_a\to 1-\lambda_a$. A fully filled ($\lambda_a=1$) and a fully empty ($\lambda_a=0$) intracell mode are equally invisible to $S_{\rm cell}$, so it discards the inert core and counts only the shared modes.

The cell entropy is lower bounded by an average quantum distance of the states, that is 
\begin{equation}
S_{\rm cell}\ge 4\ln 2\int_{\rm BZ}
\frac{d^dk}{(2\pi)^d}\frac{d^dk'}{(2\pi)^d}
\big(1-|\langle u_\bk|u_{\bk'}\rangle|^2\big)
\label{eq:Scell_bound_mt}
\end{equation}
(see Appendix~\ref{app:entropy-cumulants}).
The right-hand side captures the average of quantum distance between the wavefucntions and vanishes only when all wavefunctions are
parallel across the Brillouin zone. 
Since a Chern band cannot have globally parallel states, the average distance is strictly positive and $S_{\rm cell}>0$.

The entropy $S_\mathrm{cell}$ thus provides a real-space consequence of band topology.
This is a necessary real-space consequence
of the topological obstruction.
As seen in Fig.~\ref{fighaldane_model}, $S_\mathrm{cell}$ is finite in the
topological phase and becomes exponentially small on approaching the atomic limit in the trivial phase of the Haldane model.
The derivative of the entropy with the tuning parameter peaks at the transition,
consistent with the presence of a Dirac theory at the second-order transition, where critical
fluctuations produce entanglement across all length scales \cite{Ryu2006, Fidkowski2010,Turner2010Entanglement}. Details about the model and parameters are outlined in Appendix~\ref{app:Haldane_model}.
Similar entanglement features also arise from other topological obstructions, especially when crystalline symmetries are added to the classification \cite{PaperCNO1}.

\section{Density Form Factors}
\label{secformfactors}

While the uc-RDM is an interesting object in its own right \cite{PaperCNO1}, the CNOs obtained lead to a singular value decomposition of the density form factors which is useful for representing interactions.

\subsection{Form factor SVD}
The central object sits at the intersection of two
projectors.
The projector onto the selected bands,
\begin{equation}
P
=\sum_{\bk,n\in P}
|\psi_{\bk,n}\rangle\langle\psi_{\bk,n}| \label{eq:def_P_L}
\end{equation}
where $|\psi_{\bk, n}\rangle$ are the Bloch states for band $n$ and crystal momentum $k$,
and the local projector onto the reference unit cell,
\begin{equation}
\Pi_\bR
=
\sum_{\alpha\in\mathrm{u.c.}}
|\bR,\alpha\rangle\langle \bR,\alpha | \label{eq:def_P_0}
\end{equation}
where $\alpha$ labels the degree of freedom inside the unit cell labeled by $\bR$ in real space. 
We set $\Pi = \Pi_{\bR=0}$ as the reference unit cell and consider the rectangular operator
\begin{equation}
\mathcal{U} = \Pi\,P
\end{equation}
that maps the low-energy subspace to the local unit-cell Hilbert space.
We then define two adjoint products
\begin{equation}
\Lambda = \mathcal{U}^\dagger\mathcal{U}
= P\Pi P,
\quad
L = \mathcal{U}\mathcal{U}^\dagger
= \Pi P\Pi. \label{eq:L_Lambda_relation}
\end{equation}
The second is the uc-RDM derived in Sec.~\ref{sec1pRDM}. The first, 
$\Lambda$, is the projection of the unit-cell density operator onto 
the low-energy subspace, with matrix elements
\begin{equation}
\Lambda_{\bk m, \bk' n} = \frac{1}{N_k}\braket{u_{\bk,m}}{u_{\bk',n}},
\label{eqLambda_kernel}
\end{equation}
where $|u_{\bk,n}\rangle$ are cell-periodic Bloch states in the periodic 
gauge.
The factor of $1/N_k$ follows from the normalization of the Bloch states. 
Since $|\alpha\rangle$ is supported in the reference cell, 
$\langle\alpha|\psi_{\bk,n}\rangle = \langle\alpha|u_{\bk,n}\rangle/\sqrt{N_k}$, 
and the two such overlaps entering 
$\langle\psi_{\bk,m}|\Pi|\psi_{\bk',n}\rangle$ together produce the 
$1/N_k$.
The connection between the uc-RDM and $\Lambda$ follows from the SVD
of the operator $\mathcal{U}$. Because $L=\mathcal{U}\mathcal{U}^\dagger$ and $\Lambda=\mathcal{U}^\dagger \mathcal{U}$ are the two adjoint products of one operator, they share the same nonzero spectrum $\{\lambda_a\}$. This identify offers an immense numerical advantage. The eigenvalues of the small $N_\alpha\times N_\alpha$ real-space matrix $L$ are the eigenvalues of the much larger momentum-space form-factor kernel $\Lambda$.
Diagonalizing $L$ yields eigenvalues $\{\lambda_a\}$ and
eigenvectors $\{|\tau_a\rangle\}$ (the CNOs), which are precisely the singular values squared and left singular vectors of $\mathcal{U}$. The
right singular vectors, obtained from
$\Lambda = \mathcal{U}^\dagger\mathcal{U}$, are the momentum-space envelope
functions
\begin{equation}
    |s_{a}\rangle = \frac{1}{\sqrt{\lambda_a}} \mathcal{U}^\dagger 
    |\tau_a\rangle = \dfrac{1}{N_k}\sum\limits_\bk s_{a,n,\bk} |\psi_{\bk,n} \rangle    
\label{eq:envelope_def}
\end{equation}
with $s_{a,n,\bk} = \braket{u_{\bk,n}}{\tau_a}/\sqrt{\lambda_a}$ being the associated envelope function.
These satisfy orthonormality in the discrete Brillouin-zone inner product
\begin{equation}
\frac{1}{N_k} \sum_{\bk} \sum_{n \in P} s_{a,n, \bk} \, s^*_{b,n, \bk} = \delta_{ab}.
\label{eqenvelope_orthonormality}
\end{equation}
The SVD of $\mathcal{U}$ provides the decomposition of $\Lambda$
\begin{equation}
\Lambda_{\bk m, \bk' n}  = \frac{1}{N_k}\sum_a \lambda_a \, s_{a,m, \bk} \, s^*_{a,n, \bk'},
\label{eqLambda_SVD}
\end{equation}
expressing the overlap kernel as a sum over orthogonal CNO channels, each weighted by its occupation $\lambda_a$.

The kernel $\Lambda$ is closely related, but not identical, to the density form factors appearing in projected interactions $\langle \psi_{\bk,m}| e^{ i \bq \cdot \br } | \psi_{\bk+\bq,m}\rangle$.
Unlike $\Lambda$, these depend on an unrestricted momentum transfer $\bq$ and do not involve the cell-periodic states in the periodic embedding. 
Consequently, our decomposition is not the exact SVD of the physical form-factor matrix. The advantage of working with $\Lambda$ is that its SVD is obtained by diagonalizing the much smaller uc-RDM $L$. 

\subsection{Topologically Protected Zeros}\label{subseczerostopology}

For topological bands, the envelope functions $s_{a,n,\bk}$ exhibit
protected zeros, momenta where a particular CNO mode decouples
from the Bloch wavefunction. In the
closely related language of singular connections \cite{Mera2021, Mera2021a, Mera2022}, the zeros and the associated singular gauge carry the quantum geometry.
Zeros of overlaps with fixed
trial orbitals, and the associated failure of projection-based Wannier constructions in topological bands, are already known \cite{Kohmoto1985, ThonhauserPRB2006, SoluyanovPRB2012}.

Following the periodic embedding defined in the previous section, the projection operator $P_\bk = |u_{\bk}\rangle\langle u_\bk| $ is a bundle over the Brillouin zone torus $\mathbb{T}^2$.
The leading CNO $|\tau_1\rangle$ is a fixed vector in the unit-cell Hilbert space. Its fiberwise projection $P_\bk |\tau_1 \rangle =\sqrt{\lambda_1} s_{1, \bk} |u_\bk\rangle $ is a globally defined smooth section of the Bloch line bundle over $\mathbb{T}^2$.
For a Chern band with $\mathcal{C} \neq 0$, a section of this line bundle cannot be nowhere vanishing \cite{BrouderPRL2007,Monaco2018}; its zeros carry a total winding fixed by the Chern number \cite{Thouless1982TKNN,Kohmoto1985}. By the Poincar\'e-Hopf count, $s_{1,\bk}$ must vanish at a set of points in the Brillouin zone whose vorticities sum to $\mathcal{C}$. 

\begin{figure}
    \centering
    \includegraphics[width=\columnwidth]{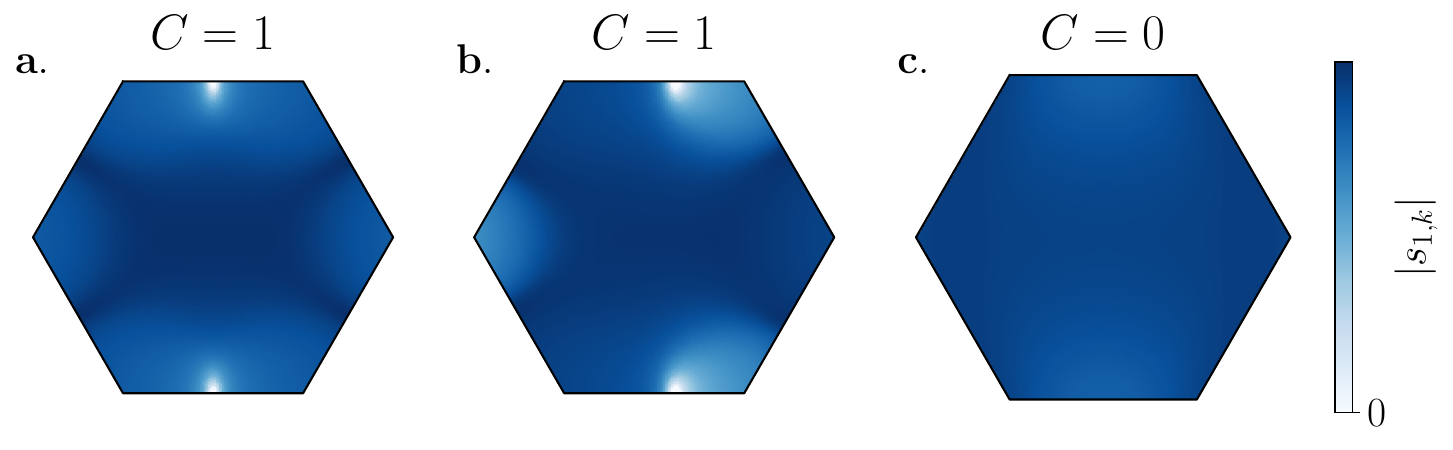}
    \caption{\textbf{Topologically protected zeros in the Haldane model.}
    Magnitude of the leading CNO envelope $|s_{1,\bk}|$ in
    (a)-(b) the topological phase and (c) the trivial phase.
    In the topological phase $s_{1,\bk}$ vanishes at a point in the
    Brillouin zone, reflecting the obstruction to a
    single localized orbital. The location of the zero depends on parameters (see Appendix~\ref{app:Haldane_model}), but the existence is a topological property. There are no zeros in the trivial phase.}
    \label{fighaldane_zeros}
\end{figure}

Individual zeros can move and annihilate in pairs of opposite vorticity, but the total must remain fixed~\cite{Mera2022}.
This forces that a single CNO cannot represent the Bloch state everywhere.
This is explicitly illustrated in 
Fig.~\ref{fighaldane_zeros} via the topological phase of the Haldane 
model. The mode envelope $|s_{1,\bk}|$ exhibits a zero whose location depends on Hamiltonian parameters and unit-cell choice, but whose existence is guaranteed by topology. 
In the trivial phase no such zero exists.

The protected zero has consequences for the CNO spectrum. More directly,
\begin{equation}
    \lambda_1 = \dfrac{1}{N_k} \sum\limits_{\bk}|\langle u_{\bk}|\tau_1\rangle|^2 \leq 1
\end{equation}
with equality if and only if $|u_\bk\rangle = e^{i\theta_\bk} |\tau_1\rangle$ for every $\bk$,
namely if the band is a single $\bk$-independent orbital and
has $\mathcal{C} = 0$.
The zero at $\bk_0$ removes weight from this sum, it implies a strict inequality
$\lambda_1 < 1$.
Together with Eq.~\eqref{eqS_cell}, $\lambda_1<1$ gives entropy $S_\mathrm{cell}>0$. 
As a result, a Chern band cannot be disentangled even at the level of a single unit cell. 
We note that the inequality is an existence statement and does not by itself bound how large the entanglement can be, as that depends on other geometric objects \cite{Budich2013, Paul2024, Kruchkov2025}.

\subsection{Optimal Truncation and Error Bounds}\label{subsec:optimal_truncation}

The CNO decomposition in Eq.~\eqref{eqLambda_SVD} hints at a low-rank approximation to the overlap kernel. Retaining the leading $N_\tau$ modes gives
\begin{equation}
\Lambda^{(N_\tau)}_{\bk m,\bk' n}
=
\dfrac{1}{N_k} \sum_{a=1}^{N_\tau}
\lambda_a \,
s_{a,m,\bk}
s^*_{a,n,\bk'},
\label{eqLambda_trunc}
\end{equation}
which minimizes the Hilbert--Schmidt error
\begin{align}
\mathcal E^{(N_\tau)}
=
\frac{1}{N_k^2}
\sum_{\bk,\bk'}
\sum_{m,n\in P}
\left|
\Lambda_{\bk m,\bk' n}
-
\Lambda^{(N_\tau)}_{\bk m,\bk' n}
\right|^2
\label{eqerror_trunc}
\end{align}
among all rank-$N_\tau$ approximations, by the Eckart-Young theorem.
The error is given exactly by $\mathcal{E}^{(N_\tau)} = \sum_{a > N_\tau} \lambda_a^2$, 
as derived in Appendix~\ref{app:error_proof}.

The truncated kernel can be realized as overlaps of auxiliary states
\begin{equation}
|\tilde u_{\bk,n}\rangle
=
\sum_{a=1}^{N_\tau}
\sqrt{\lambda_a}\,
s^*_{a,n,\bk}\,
|\tau_a\rangle,
\label{equtilde}
\end{equation}
which satisfy
$\langle \tilde u_{\bk,m} | \tilde u_{\bk',n} \rangle
=
N_k \Lambda^{(N_\tau)}_{\bk m,\bk' n}$.
However, the states $|\tilde u_{\bk,n}\rangle$ need not form an orthonormal set. In particular, zeros of the envelope functions $s_{a,n,\bk}$ will obstruct orthonormalization in topologically nontrivial bands. We return to this issue in Sec.~\ref{secCNObasis}.

\subsection{Complexity Measures}
\label{subsec:complexity_measures}
Independent of how the CNOs are employed, the error in Eq.~\eqref{eqerror_trunc} depends only on the discarded eigenvalues.
Rapid spectral decay implies that few CNOs capture the dominant correlations of the low-energy manifold, whereas slow decay signals pronounced multi-orbital character.
The CNO spectrum therefore provides a measure of the orbital complexity of the low-energy manifold, extending characterization of bands beyond standard quantum geometric quantities.
This motivates us to introduce two measures that provide quick estimates of orbital complexity.
Once the occupation numbers are normalized, $\sum_{a=1}^{N_{\rm orb}} \lambda_a = 1$ where $N_{\mathrm{orb}}$ is the number of orbitals per unit cell, they define a probability distribution over CNO modes. 
The inverse participation ratio (IPR) and normalized Shannon entropy quantify the concentration versus spread of this distribution
\begin{equation}
\mathcal{I} = \sum_{a} \lambda_a^2, 
\quad  
\mathcal{S} = -\dfrac{1}{\log N_{\mathrm{orb}} }\sum_{a} \lambda_a \log \lambda_a.
\label{eq:IPR_Shannon}
\end{equation}
The IPR satisfies $1/N_{\mathrm{orb}} \leq \mathcal{I} \leq 1$, approaching unity when a single mode dominates ($\lambda_1 \approx 1, \lambda_{a>1} \approx 0$) and $1/N_{\mathrm{orb}}$ when all modes contribute equally. The normalized Shannon entropy $\mathcal{S}$ ranges from $0$ (single mode) to $1$ (uniform distribution).

In the CNO framework, the IPR characterizes the truncation error in Eq.~\eqref{eqerror_trunc}. 
A large IPR ($\mathcal{I}\to 1$) indicates that the state is concentrated 
in only a few modes, so the truncation error remains small even at low rank, 
whereas a small IPR ($\mathcal{I}\to 1/N_{\mathrm{orb}}$) indicates that 
the weight is distributed over many modes and therefore more modes are 
required for accurate reconstruction. Consequently, for a fixed target error $\mathcal{E}_{\mathrm{target}}$, the IPR provides an estimate of the minimal truncation rank required.

The Shannon entropy $\mathcal{S}$ can now be interpreted in terms of unit-cell entanglement $S_{\mathrm{cell}}$. The two measures coincide up to normalization and corrections from the $(1-\lambda_a)\ln(1-\lambda_a)$ terms in $S_{\mathrm{cell}}$. 
Both quantify how strongly the unit-cell state entangles different orbital degrees of freedom.

\begin{figure}
    \centering
    \includegraphics[width=\columnwidth]{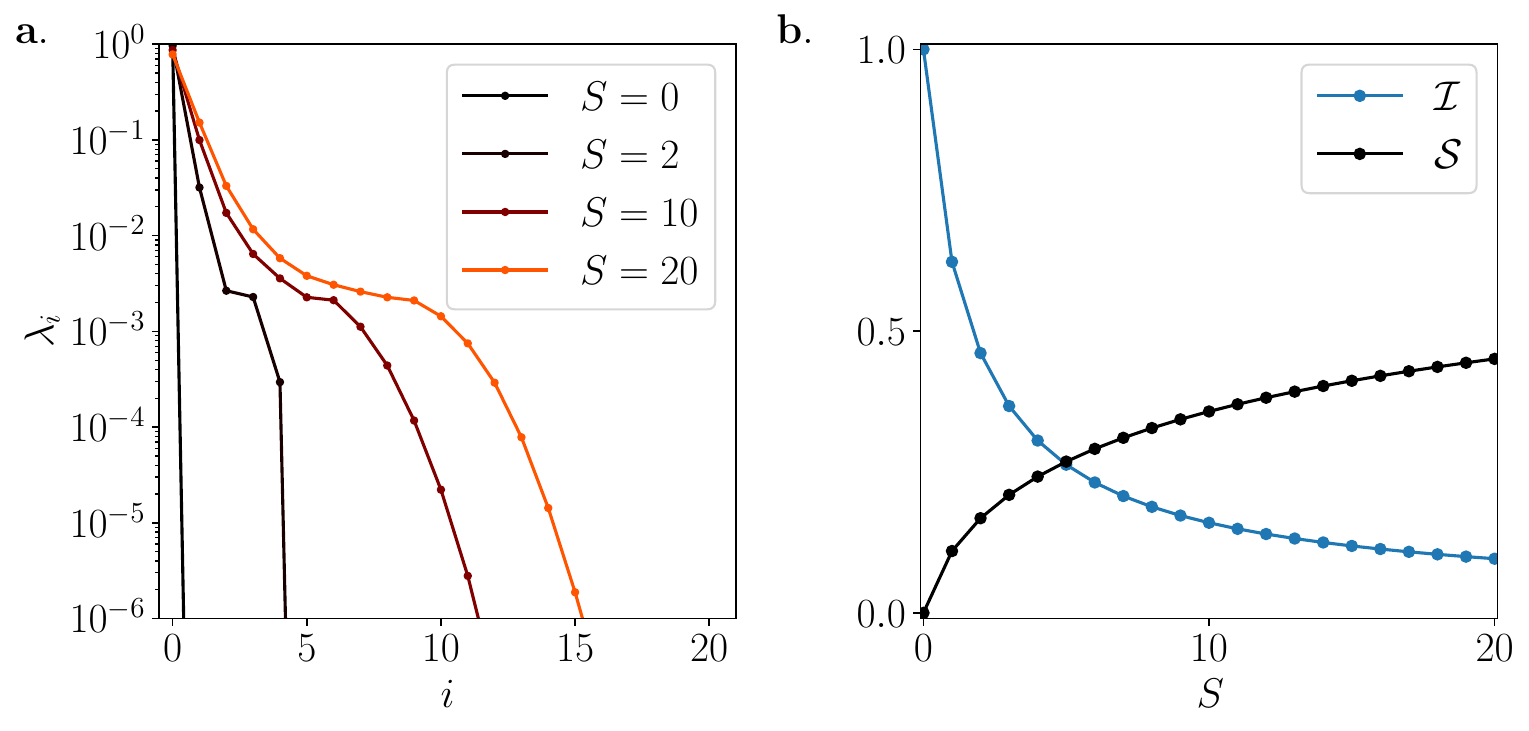}
    \caption{\textbf{Complexity measures for the multifold fermion model.}
    Continuum model $H_\bk = k_x\sigma_x+k_y\sigma_y+(k^2/2M+1)\sigma_z$
    in the $(2S+1)$-dimensional representation of $SU(2)$, analyzing the
    top band.
    (a) Occupation spectrum for increasing spin $S$, showing growth in
    the number of significant modes.
    (b) IPR $\mathcal{I}$ decreases as more modes contribute; Shannon
    entropy $\mathcal{S}$ increases as spectral weight spreads.
    The model interpolates from the single-orbital limit
    ($S=0$: $\mathcal{I}=1$, $\mathcal{S}=0$) to the ideal-band regime
    ($S\to\infty$: $\mathcal{I}\to 0$, $\mathcal{S}\to 1$), with the number of significantly occupied modes growing with $S$.
    Momentum cutoff $\Lambda_K\gg M$ held fixed.}
    \label{fig:multifold}
\end{figure}

To illustrate these measures in a model, we consider a multifold fermion defined in Appendix \ref{app:MultiFold_fermion}.
Unlike tight-binding models where the Brillouin zone provides a natural momentum cutoff, this model require an explicit choice. 
The momentum cutoff $\Lambda_K$ defines the spatial extent $\sim 1/\Lambda_K$ of the real-space partition. 
It determines which length scale acts as the "unit cell" for constructing $L$. For the multifold model, we fix $\Lambda_K \gg M$ and vary $S$, ensuring the cutoff encompasses the region where band inversions modify wavefunction structure.

Fig.~\ref{fig:multifold} shows the CNO spectrum and complexity measures as $S$ increases. At $S=0$ (trivial band), a single CNO suffices with $\mathcal{I}=1$, $\mathcal{S}=0$, and rank-1 truncation is exact. As $S$ grows, the number of significant modes increases, reflecting the multiple band inversions required for high Chern number. The IPR decreases and Shannon entropy increases, quantifying the growing orbital complexity. By $S=3$ (Chern number $|\mathcal{C}|=6$), approximately 7 modes carry significant weight. Capturing the spectral weight of the  band to small truncation error $\mathcal{E}^{(N_\tau)}$ requires a number of modes that grows with $S$, as the increasingly ideal quantum
geometry spreads weight across many CNO channels.

\section{CNOs for projected interactions}
\label{secCNObasis}
CNOs are local orbitals extracted from the low-energy band wavefunctions. 
We shall use these orbitals to reveal a low-rank structure in projected interactions. 
We will see that the CNO occupation spectrum $\{\lambda_a\}$ determines both the minimal number of orbitals required to approximate the density operator to a desired accuracy and the resulting interaction scales.

\subsection{Construction of Truncated Wavefunctions}\label{subsec:truncatedWF}

\begin{figure}
    \centering
    \includegraphics[width=\columnwidth]{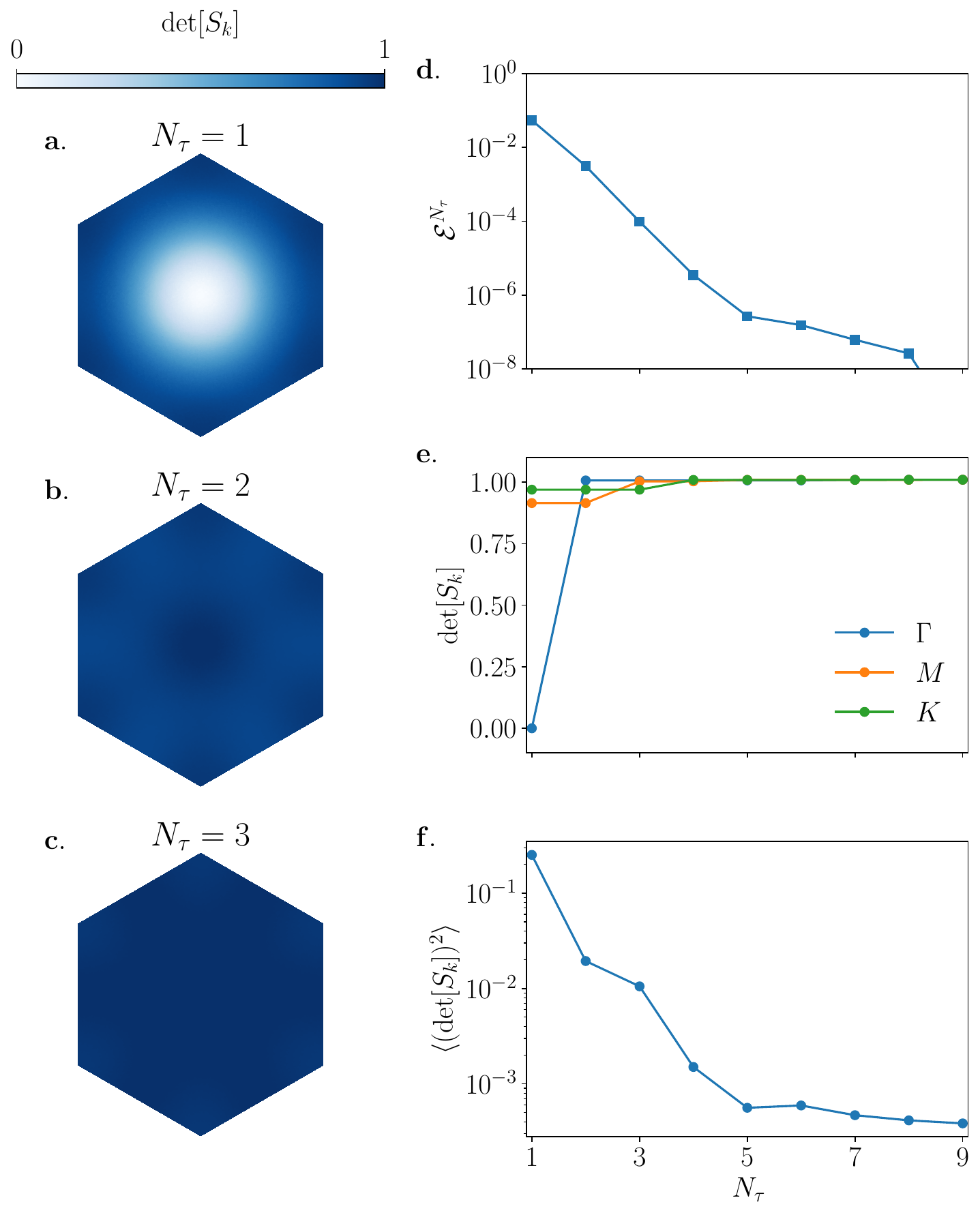}
    \caption{\textbf{Gram matrix in chiral TBG.}
    Determinant of the Gram matrix $\mathcal{S}_{\bk}$,
    Eq.~\eqref{eqSk}, over the {\moire} Brillouin zone for truncations
    $N_\tau = 1$--$3$ (top panels), for the TBG flat band at the magic angle in the chiral limit. For $N_\tau = 1$ the matrix is singular at
    the topological zero at $\Gamma$; the zero is lifted for $N_\tau \ge 2$.
    The bottom panel shows the truncation error $\mathcal{E}^{(N_\tau)}
    = \sum_{a > N_\tau} \lambda_a^2$ on a log-linear scale, determinant of the Gram matrix at the high symmetry points, and its variance as a function of $N_\tau$.}
    \label{fig:gram_conditioning}
\end{figure}

Given the CNOs $\{|\tau_a\rangle\}$ of the uc-RDM, we introduce the
fermionic operator $c^\dag_{\bR,a}$ that creates the state $|\tau_a\rangle$
in unit cell $\bR$, with Bloch transform
\begin{equation}
    c^\phdag_{\bk, a} = \frac{1}{\sqrt{N_k}} \sum_{\bR} e^{-i \bk\cdot \bR}\,
    c^\phdag_{\bR,a},
\end{equation}
where $N_k$ is the number of unit cells. The envelope functions $s_{a,n,\bk}$
defined in Eq.~\eqref{eq:envelope_def} give the exact decomposition of the
low-energy Bloch operators in the CNOs,
\begin{equation}
    c^\phdag_{\bk, n} = \sum_{a} \sqrt{\lambda_a}\, s^*_{a, n, \bk}\,
    c^\phdag_{\bk, a},
    \label{eq:bloch_exact}
\end{equation}
which follows from the completeness of the CNOs within the
low-energy subspace. Retaining only the $N_\tau$ CNOs with the largest occupations $\lambda_a$ defines the truncated Bloch operators
\begin{equation}
    \tilde{c}^\phdag_{\bk, n} = \sum_{a=1}^{N_\tau} \sqrt{\lambda_a}\,
    s^*_{a, n, \bk}\, c^\phdag_{\bk, a}
    \label{eq:bloch_truncated}
\end{equation}
which is identical to Eq.~\eqref{equtilde} but written in terms of operators.
As stated before, the associated cell-periodic states provide
the optimal rank-$N_\tau$ approximation to the overlap kernel $\Lambda_{\bk m,\bk'n}$, but are not generally orthonormal. Their overlaps
define the Gram matrix
\begin{equation}
[\mathcal{S}_\bk]_{m,n} = \langle \tilde{u}_{\bk,m} | \tilde{u}_{\bk,n}
\rangle = \sum_{a=1}^{N_\tau} \lambda_a \, s_{a,m,\bk}\, s^*_{a,n,\bk},
\label{eqSk}
\end{equation}
Next, following standard L\"{o}wdin orthogonalization yields the orthonormal operators
\begin{equation}
\bar{c}^\phdag_{\bk, n} = \sum_{m} [\mathcal{S}_\bk^{-1/2}]_{n,m} \,
\tilde{c}^\phdag_{\bk, m}.
\label{equbar}
\end{equation}
At momenta where topologically protected zeros of $\mathcal{S}_\bk$ occur,
the orthonormalization breaks down. This is a direct manifestation
of the fact that topological bands cannot be represented by a single local orbital. 
The number of additional orbitals required to regularize
$\mathcal{S}_\bk$ depends on whether the zeros of the leading envelope are coincident or separated in the Brillouin zone.

When all zeros of $s_{1,n,\bk}$ are coincident at a single momentum
$\bk_0$, completeness of the CNOs,
$\sum_a \lambda_a|s_{a,n,\bk}|^2 = 1$, guarantees the
existence of at least one subleading mode $b$ with
$|s_{b,n,\bk_0}| > 0$. 
A single additional orbital will therefore be able to span the entire Brillouin zone, make the norm in Eq.~\eqref{eqSk} finite and furnish a two-orbital model for any given Chern number.
On the other hand, when the zeros of $s_{1,n,\bk}$ are separated, sitting at distinct momenta $\bk_1, \bk_2, \ldots$, the situation is more subtle.
Completeness ensures that each $\bk_i$ is covered by at least one nonvanishing mode, but different zeros may require different modes.
Nevertheless, since all bands with the same Chern number belong to the same topological class, there should exist a well-conditioned transformation mapping a band with multiple zeros to one with a single zero of higher vorticity. In that representation, a two-orbital model should again suffice for arbitrary Chern number.
The above argument ignores symmetries which can further enrich this classification \cite{PaperCNO1}.

While the CNOs can be used to construct effective Bloch Hamiltonians, there is a fundamental problem that comes with CNO representation.
As we saw in the example of Chern band, one CNO is unable to capture all the states and one must invoke at least two CNOs. 
If one were to use the upfolded representation, the effective model will contains more orbitals than target bands. Consequently, there will be zero modes that reflect the redundancy of the CNOs. 
If the objective is to obtain a tight-binding model, the CNOs instead provide an excellent set of trial orbitals for Wannierization, an approach that is explored in Ref.~\cite{PaperCNO1} for various {\moire} materials.
In the present work we instead use the CNOs to simplify projected interactions, where it naturally reveals a low-rank structure in the density operators.

\subsection{Projected Interactions and CNO Channels}
\label{subsec:projected_interactions}

\begin{figure*}
    \centering
    \includegraphics[width=1.8\columnwidth]{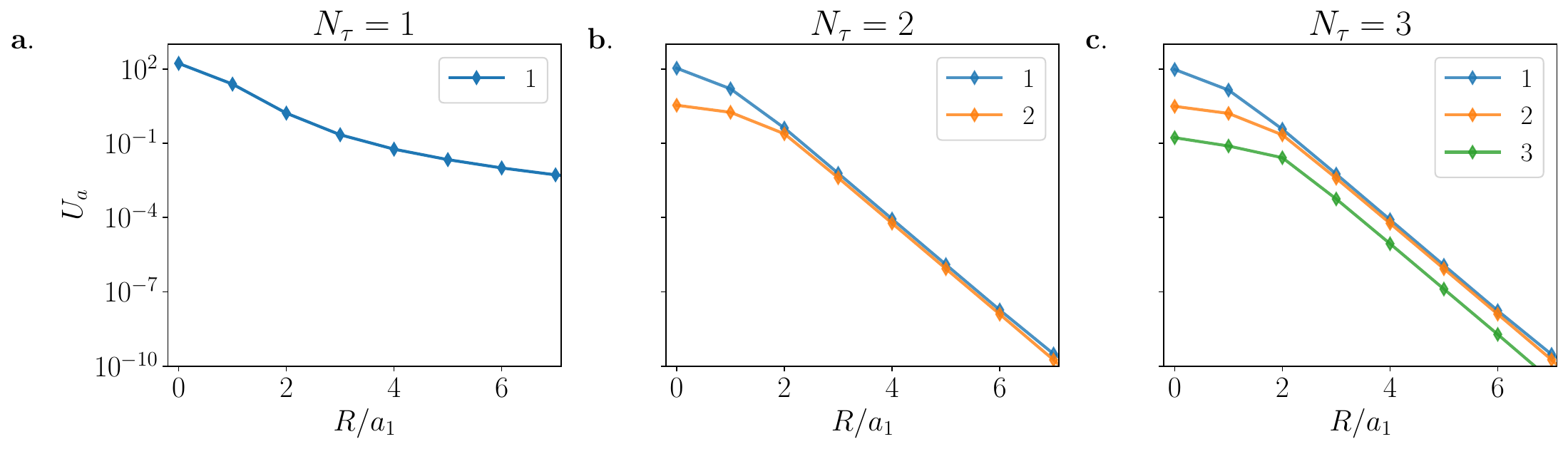}
    \caption{\textbf{Real-space interactions for chiral TBG.}
Density-density interaction strengths $U_a(\bR=n \ba_1)$ for a screened Coulomb interaction $V_\bq=V_0\tanh(q\xi)/(q\xi)$ with screening length $\xi=10\,\mathrm{nm}$. All plots use a log-linear scale. (a) For $N\tau=1$, the interaction exhibits a power-law profile owing to topological obstruction. (b) Including the obstruction-covering CNO ($N_\tau=2$) regularizes the truncated Gram matrix $\mathcal{S}_{\mathbf{k}}$, removing its singularity and producing interactions that decay exponentially with distance. The additional orbital preserves the hierarchy of on-site interactions ($R=0$).
}
    \label{fig:range_of_interaction}
\end{figure*}

To make analytic progress, we consider a single energetically isolated band
$n$ (per spin). In this case the Gram matrix in Eq.~\eqref{eqSk} reduces
to a scalar, and we assume a truncation to $N_\tau$ orbitals.

Following the standard prescription~\cite{Kwan2025_moire_review}, projecting a UV density-density interaction onto the low-energy manifold
yields
\begin{equation}
    \bar{H}_{\rm int} = \frac{1}{2A} \sum_{\sigma \sigma'} \sum_\bq V_\bq\,
    \bar{\rho}_{\bq,\sigma}\, \bar{\rho}_{-\bq,\sigma'},
    \label{eq:proj_int_mt}
\end{equation}
where $A$ is the system area, $V_\bq$ is the Fourier transform of the UV
interaction, and
\begin{equation}
    \bar{\rho}_{\bq,\sigma} =
    \sum_{\bk \in {\rm BZ}}
    \Lambda_{\bk,\bk+\bq,\sigma}^{\rm phys}\,
    c^\dag_{\bk, n, \sigma}\,
    c^\phdag_{\bk+\bq, n, \sigma} \label{eq:bar_rho}
\end{equation}
is the projected density operator. 
Leaving the derivation to Appendix~\ref{app:projectedInteractions}, we note here that $\Lambda_{\bk,\bk+\bq,\sigma}^{\rm phys}$ is the form factor that uses the cell-periodic states $\Lambda^{\rm phys}_{\bk,\bk+\bq,\sigma} = \langle u^{\rm phys}_{\bk,n,\sigma} | u^{\rm phys}_{\bk+\bq,n,\sigma}\rangle/N_k$, that are not periodic in general $| u^{\rm phys}_{\bk+\bG,n,\sigma}\rangle \neq | u^{\rm phys}_{\bk,n,\sigma}\rangle$ where $\bG$ is a reciprocal lattice vector. 
Moreover, the sum over $\bq$ is unbounded.

Although the physical form factor in Eq.~\eqref{eq:bar_rho}
is distinct from the kernel diagonalized in Eq.~\eqref{eqLambda_SVD}, the CNO decomposition nevertheless induces a
strong hierarchy in the band projected interaction matrix elements. 
The origin of this hierarchy is not obvious a priori, however as we show below, it works well for magic-angle twisted bilayer graphene in the chiral limit.

Substituting the truncated expansion Eq.~\eqref{eq:bloch_truncated} gives
\begin{equation}
    c_{\bk, n, \sigma} =
    \frac{1}{\sqrt{N_k}}
    \sum_{\bR}\, e^{-i \bk\cdot \bR}
    \sum_{a=1}^{N_\tau} \dfrac{\sqrt{\lambda_a}\,
    s_{a,n,\bk}}{\sqrt{\mathcal{S}_\bk}}\,
    c_{\bR, a, \sigma},
    \label{eq:bloch_cno_decomp}
\end{equation}
and the projected density operator becomes
\begin{equation}
    \bar{\rho}_{q,\sigma} =
    \sum_{\bR_1, \bR_2}
    \sum_{a,b=1}^{N_\tau}
    \mathcal{F}_{a,b}^\sigma(\bq, \bR_1, \bR_2)\,
    c^\dag_{\bR_1, a, \sigma}\,
    c^\phdag_{\bR_2, b, \sigma},
\end{equation}
where
\begin{align}
    \mathcal{F}_{a,b}^\sigma(\bq, \bR_1, \bR_2)
    &=
    \frac{\sqrt{\lambda_a \lambda_b}\; e^{-i \bq\cdot \bR_2}}{N_k}
    \sum_\bk\,
    e^{i \bk\cdot(\bR_1 - \bR_2)} \nonumber \\
    &\phantom{=}\hspace{1.0cm}
    \times\,
    \Lambda_{\bk,\bk+\bq,\sigma}^{\rm phys}\,
    \dfrac{s^*_{a,n,\bk}\, s_{b,n,\bk+\bq}}{ \sqrt{ \mathcal{S}_\bk \mathcal{S}_{\bk+\bq} } }.
    \label{eq:Fab}
\end{align}
The band projected interaction is thus generically nonlocal in real space CNO basis,
\begin{align}
    \bar{H}_{\rm int}
    &=
    \sum_{\substack{\{\bR_i\},\{a_i\} \\ \sigma,\sigma'}}
    V_{ \{ \bR_i \} }^{ \{a_i\} }\,
    c^\dag_{\bR_1, a_1, \sigma}
    c^\phdag_{\bR_2, a_2, \sigma}
    c^\dag_{\bR_3, a_3, \sigma'}
    c^\phdag_{\bR_4, a_4, \sigma'}.
\end{align}
The nonlocality is a generic consequence of projecting interactions onto topological bands. For a single Wannier function,
topology appears through power-law tails in interaction. In this scheme of normalized CNOs (defined in Eq.~\eqref{eq:bloch_cno_decomp}),
the same obstruction is absorbed into additional local orbitals, leading to a hierarchy of predominantly onsite interactions. Increasing the
number of retained CNOs therefore trades orbital number for locality.

Defining the onsite form factor $F_{ab}^\sigma(\bq) \equiv
\mathcal{F}_{ab}^\sigma(\bq,0,0)$,
\begin{equation}
    F_{ab}^\sigma(\bq) =
    \frac{\sqrt{\lambda_a \lambda_b}}{N_k}
    \sum_{\bk}\,
    \Lambda_{\bk,\bk+\bq,\sigma}^{\rm phys}\,
    \dfrac{ s^*_{a,n,\bk}\, s_{b,n,\bk+\bq}}{ \sqrt{ \mathcal{S}_\bk \mathcal{S}_{\bk + \bq} } },
    \label{eq:Fab_onsite}
\end{equation}
the intra-orbital Hubbard interaction follows from setting $a_i = a$ and
$\bR_i = \bR$ for all $i$,
\begin{equation}
    U_a =
    \dfrac{1}{2A}
    \sum_\bq
    V_\bq\,
    F_{aa}^{\uparrow}(\bq)\,
    F_{aa}^{\downarrow}(-\bq),
\end{equation}
and the onsite inter-orbital interaction is
\begin{equation}
    V_{ab} =
    \frac{1}{2A}
    \sum_\bq
    V_\bq\,
    F_{aa}^{\uparrow}(\bq)\,
    F_{bb}^{\downarrow}(-\bq),
\end{equation}
with analogous expressions for other inter-orbital channels.

Fig.~\ref{fig:range_of_interaction} shows the interaction strengths for chiral TBG at
the magic angle using
$V_\bq=V_0\tanh(q\xi)/(q\xi)$ with $\xi=10\,\mathrm{nm}$. 
We note two striking features in the projected interactions. The first is that, despite the physical form factors being unrelated to CNO via diagonalization, a hierarchy is clearly visible at on-site interactions ($R=0$ in the three panels). 
The dominant interaction
is the onsite Hubbard term of the leading CNO, with both higher CNO
channels and nonlocal interactions progressively suppressed.
The second feature is the absence of power-law tails in projected interaction once $N_\tau \geq 2$.

The robustness of the CNO hierarchy in the on-site interaction is tied to the concentration of charge density in the flat-band manifold near the $AA$ sites. This concentration leads to a large weight $\lambda_1$ for the first CNO. The second CNO captures states near the $K$ points (see Fig.~\ref{fig:chiral_msi}), which constitute a smaller portion of the low-energy manifold. An orbital with larger spectral weight has a larger CNO occupation (in the sense of single-particle entanglement), is more localized (see Fig.~\ref{fig1}{\bf c}) and therefore incurs a larger energetic cost when doubly occupied.

The vanishing of power tails can be understood from a cancellation of phase singularity between the form factor $\Lambda^{\rm phys}$ and the CNO envelopes $s_{a,n,\bk}$ in the density structure factor $F_{aa}(q)$. Starting from Eq.~\eqref{eq:Fab_onsite} of, we first introduce $C_{a,n,\bk} = s_{a,n,\bk} /\sqrt{\mathcal{S}_{\bk}}$ to rewrite the structure factor as
\begin{equation}
F_{ab}^\sigma(q) =
    \frac{\sqrt{\lambda_a \lambda_b}}{N_k}
    \sum_{\bk}\,
    \Lambda_{\bk,\bk+\bq,\sigma}^{\rm phys}\,
    C^*_{a,n,\bk}\, C_{a,n,\bk+\bq}
\end{equation}
and then expand the diagonal $F_{aa}(\bq)$ for small $\bq$ to locate the topological obstruction.
The form-factor expansion gives
\begin{equation}
\Lambda^{\rm phys}_{\bk,\bk+\bq}
= 1 + i\,\bA(\bk)\!\cdot\!\bq + \cdots,
\label{eq:Lambda_expand}
\end{equation}
with $\bA$ the Berry connection $\langle u_\bk|\partial_\bk u_\bk\rangle = i\bA(\bk)$.
Second, there is a contribution from the phase of $C^*_{a,n,\bk}C_{a,n,\bk+\bq}$.
Since $\mathcal{S}_\bk$ is real and positive, it contributes no phase, and if $\phi_{a,\bk} = \arg C_{a,n,\bk} $, we get
\begin{equation}
\nabla \arg\!\big(C^*_{a,n,\bk} C_{a,n,\bk+\bq}\big)\big|_{\bq\to0}
= \nabla\phi_{a,\bk}.
\end{equation}
This gradient is singular at a zero $\bk_0$ of $s_{a,n,\bk}$, where $\phi_{a,\bk}$ winds by $2\pi$ times the vorticity of the zero.
The dependence on $\mathcal{S}_\bk$ enters only through the modulus $|C_{a,n,\bk}|^2 = |s_{a,n,\bk}|^2/\mathcal{S}_\bk$, which weights this singular gradient and ultimately the contribution is
\begin{equation}
\lambda_a\,\bq\!\cdot\!\!\int_{\bk_0} \!d\bk\;
\frac{|s_{a,n,\bk}|^2}{\mathcal{S}_\bk}\,\nabla\phi_{a,\bk}.
\label{eq:ImF_weighted}
\end{equation}
Whether the winding survives is decided by whether this weight vanishes at $\bk_0$, and hence by whether $\mathcal{S}_\bk$ remains finite.

\begin{figure}
    \centering
    \includegraphics[width=\columnwidth]{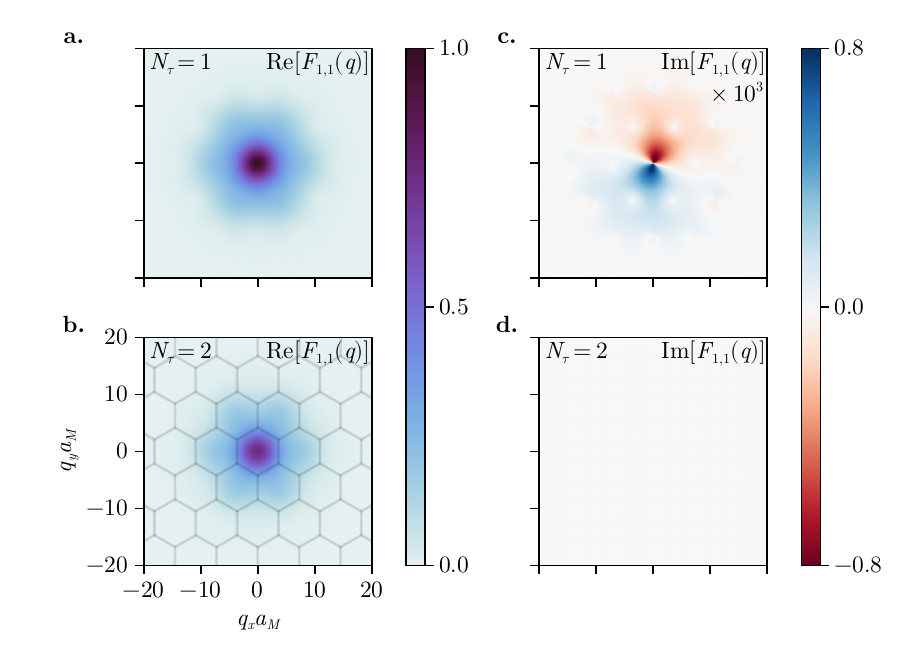}
    \caption{\textbf{Onsite structure factor $F^\sigma_{ab}(\bq)$ for the chiral TBG flat band.} (a) Real part of $F_{11}(\bq)$ for a single-CNO truncation $N_\tau=1$, with (c) its imaginary part, which shows a branch cut at small $\bq$ from the winding of $\nabla\arg s_{1,n,\bk}$ around the protected zero of the leading chiral TBG envelope. (b) Real part of $F_{11}(\bq)$ for $N_\tau=2$, with (d) the imaginary part of the diagonal structure factor, which vanishes once the subleading mode keeps $\mathcal{S}_\bk$ finite at the zero and removes the winding.}
    \label{fig:structure_factor_Fq}
\end{figure}

For $N_\tau=1$, the projected norm is $\mathcal{S}_\bk = \lambda_1|s_{1,n,\bk}|^2$, which vanishes at the $\Gamma$ point and nothing cancels the phase singularity embedded in $\Lambda^{\rm phys}$.
The winding survives, and $\mathrm{Im}\,F_{11}(\bq)$ carries a non-analytic branch cut from the origin, Fig.~\ref{fig:structure_factor_Fq}(c).

For $N_\tau=2$, the subleading envelope keeps $\mathcal{S}_{\bk_0} = \lambda_2|s_{2,n,\bk_0}|^2 > 0$ finite at the zero of $s_{1,n,\bk}$.
The winding no longer survives the $\bk$-average, and $\mathrm{Im}\,F_{aa}(\bq) = 0$ as shown in Fig.~\ref{fig:structure_factor_Fq}(d).
The finite $\mathcal{S}_{\bk_0}$ supplied by the second CNO cancels the singularity. 
The result is exponentially decaying interactions as shown in Fig.~\ref{fig:range_of_interaction}.
We note however, that the second CNO is completely delocalized in energy.
This is consistent with the anomaly of TBG which restricts a local representation for any finite set of low-energy bands \cite{TBG2}.
As shown in ref.~\cite{song2022magic}, the obstruction can only be removed by introducing additional degrees of freedom beyond the minimal low-energy subspace. In our construction, the second CNO provides precisely such an additional degree of freedom, regularizing the singularity while necessarily extending beyond a finite low-energy energy window.

\subsection{Momentum dependent correlations}
\label{subsec:momentum_decoupling}

The topologically protected zero of the leading CNO envelope has a direct consequence for interaction effects. At a momentum $\bk_0$ where $s_{1,n,\bk_0}=0$, the low-energy Bloch state has no weight on the dominant CNO. Consequently, interactions acting primarily in that channel have vanishing projected matrix elements, leading to a momentum-dependent hierarchy of correlation scales determined by the wavefunction geometry. 

To illustrate this mechanism, consider an interaction consisting of independent Hubbard terms in each CNO channel,
\begin{equation} 
H_\mathrm{int} = \sum_a U_a \sum_{\bR} n_{\bR,a,\uparrow}\, n_{\bR,a,\downarrow}. \label{eq:interactionCNO_msi} \end{equation} 
In the atomic-limit Hubbard-I approximation, the self-energy is approximately diagonal in the CNO basis and satisfies $\Sigma_a(i\omega)\propto U_a^2/i\omega$. Projecting onto the isolated band gives 
\begin{equation} \Sigma(\bk,\omega) = \sum_a \lambda_a |s_{a,n,\bk}|^2 \Sigma_a(\omega). \label{eq:sigma_projected} \end{equation} 
The projected self-energy therefore inherits the CNO hierarchy. At generic momenta it is dominated by the leading CNO and produces a correlation scale of order $U_1$, whereas at $\bk_0$ the leading contribution vanishes identically. The self-energy is then controlled by the subleading channel and is parametrically reduced when the interactions satisfy $U_2\ll U_1$. More generally, whenever the CNO spectrum is hierarchical, the leading envelope function determines the momentum dependence of the dominant correlation scale.

\begin{figure}
    \centering
    \includegraphics[width=\columnwidth]{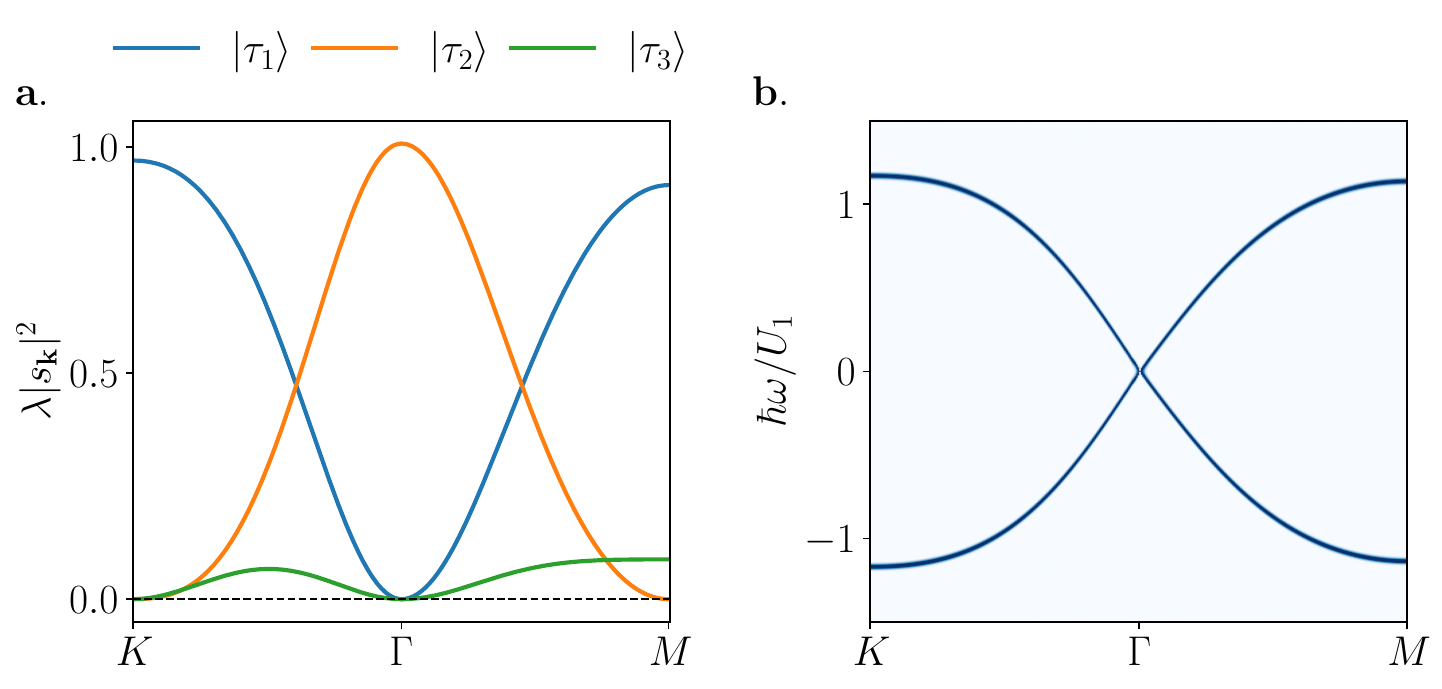}
    \caption{\textbf{Momentum-dependent correlations in chiral TBG.}
    (a)~CNO band weights $\lambda_a|s_{a,n,\bk}|^2$ along the
    $K$--$\Gamma$--$M$ path for $\theta \approx 1.07^\circ$. The band
    is carried almost entirely by the leading mode at $K$ and $M$, and by
    the subleading mode at $\Gamma$, where the leading envelope has its
    topologically protected zero.
    (b)~Single-particle spectral function $A(\bk,\omega)$ in the
    Hubbard-I approximation at half-filling where only the dominant mode develops the pole in self-energy (see Eq.~\eqref{eq:sigma_projected}. The gap is $\Delta \propto U_1$ at generic momenta and closes
    into a gapless crossing at $\Gamma$, where the leading weight
    vanishes.
    }
    \label{fig:chiral_msi}
\end{figure}

Chiral TBG provides a realization of this mechanism. The dominant CNO has a symmetry-protected zero at the $\Gamma$ point, so a Hubbard interaction acting only on the leading CNO captures the qualitative momentum dependence of the spectral function at charge neutrality, shown in Fig.~\ref{fig:chiral_msi}. The spectrum is gapped at generic momenta, where the leading CNO dominates, while the gap closes at $\Gamma$, where its weight vanishes. This single-band description is, however, not a complete theory of the Mott semimetal phase in TBG. In particular, improving the treatment of the dominant self-energy beyond Hubbard-I generically gaps the crossing within the projected model (see Appendix~\ref{app:momentum_selective_correlation}), whereas the crossing in TBG is protected by the particle-hole anomaly and therefore cannot be gapped at any order of approximation \cite{Calugaru2023}. 

The missing ingredient is that TBG contains two Chern bands per spin and valley rather than the single chiral band considered above. The CNO hierarchy therefore determines the momentum dependence of the correlation scale only within a given chiral sector, while the fate of the crossing at $\Gamma$ is governed by the coupling between the two sectors. To incorporate both sectors, we need to consider CNO states $|\tau_{a,\pm,\sigma}\rangle$, where $\pm$ labels the chirality of mode $a$. Now since particle-hole symmetry forbids mixing between opposite chiralities at $\Gamma$ \cite{Calugaru2023, ledwith2024nonlocal,song2022magic, Datta2023Heavy}, the gapless crossing reappears once both chiral sectors are included. 
A complete description therefore requires treating the full projected interaction in the CNO basis including both chiral sectors, for example within DMFT or exact diagonalization, which we leave for future work.

\section{Choice of Local Orbitals}
\label{sec:comparison}
The projected Hamiltonian and interactions derived above may be written in
any complete set of local orbitals. 
The essential question is therefore
not whether one can choose a basis, but which basis admits the most efficient truncation. This question has a long history in quantum chemistry and electronic structure \cite{Lowdin1955,Kivelson1982,RestaSorella1999,Resta2011Insulating,Knizia2013IAO,Weinhold1983NBO}, where numerous localization schemes have been developed, including Foster-Boys \cite{Foster1960}, Edmiston-Ruedenberg \cite{Edmiston1963}, Pipek-Mezey \cite{Pipek1989}, maximally localized Wannier functions \cite{Marzari1997, Marzari2012, Souza2001}, SCDM \cite{Damle2015,Damle2017}, and more
recently compact molecular orbitals \cite{Xie_cmo_1,Xie_cmo_2,Xie_cmo_3,Xie_cmo_4,Li2026Topology} and coherent states \cite{Okuma2026LocalizedBasis, li2024constraints}.
These constructions differ in their localization criteria, but all seek local orbitals motivated by the fact that a localized single orbital concentrates charge in real space and so carries a large on-site interaction.

CNOs are optimized according to a different principle. They diagonalize the
projected unit-cell density matrix and therefore provide an optimal representation of the projected Hilbert space. Their distinguishing feature is the hierarchy of occupations $\lambda_a$, which ranks the
importance of each orbital in reconstructing the low-energy Bloch states. Truncating to the leading $N_\tau$ CNOs minimizes the error in the overlap kernel, and, as shown in Sec.~\ref{subsec:projected_interactions},
organizes the dominant interaction channels by strength.
In particular, the compact molecular orbitals scheme uses a localized trial
function to seed a partial Wannierization and then construct its complementary topological orbital. It does not provide an intrinsic spectrum ranking additional modes.
By replacing the single trial function with the complete unit-cell space, the CNO is the ansatz-free completion of the projection step.

The utility of this hierarchy depends on the many-body method. Approaches such as Hartree-Fock retain the full quantum-geometric form
factors and are therefore largely insensitive to the choice of local basis. 
In contrast, methods based on local interactions, including DMFT
and more general quantum embedding schemes
\cite{Georges1996,Knizia2012}, depend explicitly on the choice of orbitals defining the impurity problem. For these methods, the CNOs provides a starting point whose interaction scales are ordered by construction. Whether this ordering coincides with the hierarchy of physical correlations is material dependent, but it offers a controlled and systematically improvable approximation for {\moire} systems.

The hierarchy of CNOs shows the tradeoff between the number of orbitals retained and their locality. For a single band,
\begin{equation}
    c_{\bk,n}
    =
    \frac{1}{\sqrt{\mathcal{S}_\bk}}
    \sum_{a=1}^{N_\tau}
    \sqrt{\lambda_a}\,
    s_{a,n,\bk}\,
    c_{\bk,a},
    \label{eq:cbar_singleband}
\end{equation}
where $\mathcal{S}_\bk=\sum_{a=1}^{N_\tau}\lambda_a|s_{a,n,\bk}|^2$ measures
how well the truncated orbitals spans the Bloch state at momentum $\bk$.
The locality of the orthonormalized orbitals is controlled by this Gram
factor. If $\mathcal{S}_\bk$ becomes small, the prefactor
$1/\sqrt{\mathcal{S}_\bk}$ develops strong momentum dependence, which
translates into long-ranged real-space tails after Fourier
transformation. Adding additional CNOs improves the coverage of the
Brillouin zone, makes $\mathcal{S}_\bk$ smooth, and produces more localized
orbitals. 

We can see this interplay precisely in the multifold model on a square lattice,
$H_\bk = \bd_\bk \cdot \bsigma$, with
\begin{equation}
\bd_\bk = (\sin k_x,\, \sin k_y,\, 1 - \cos k_x - \cos k_y),
\label{eq:lattice_multifold}
\end{equation}
where $\bsigma$ is the spin-$S$ representation of the $SU(2)$ algebra.
For $S=2$ the top band has Chern number $|\mathcal{C}|=4$, and the unit
cell carries $2S+1 = 5$ internal states, so there are five CNO modes.
The five modes span the full cell Hilbert space, so retaining all of
them is exact and the truncation error vanishes. We compare three
truncations of this complete set.

Retaining all five modes gives
\begin{equation}
\mathcal{S}_\bk = \sum_{a=1}^{5} \lambda_a|s_{a,\bk}|^2 = 1,
\end{equation}
independent of $\bk$, since the completeness of the CNOs makes the sum the norm of the full band state. The factor
$\mathcal{S}_\bk^{-1/2}$ is then a constant, the orthonormal orbitals
carry no $\bk$-dependent dressing, and the projected interaction is as
local as the model allows.

\begin{figure}
  \centering
  \includegraphics[width=\columnwidth]{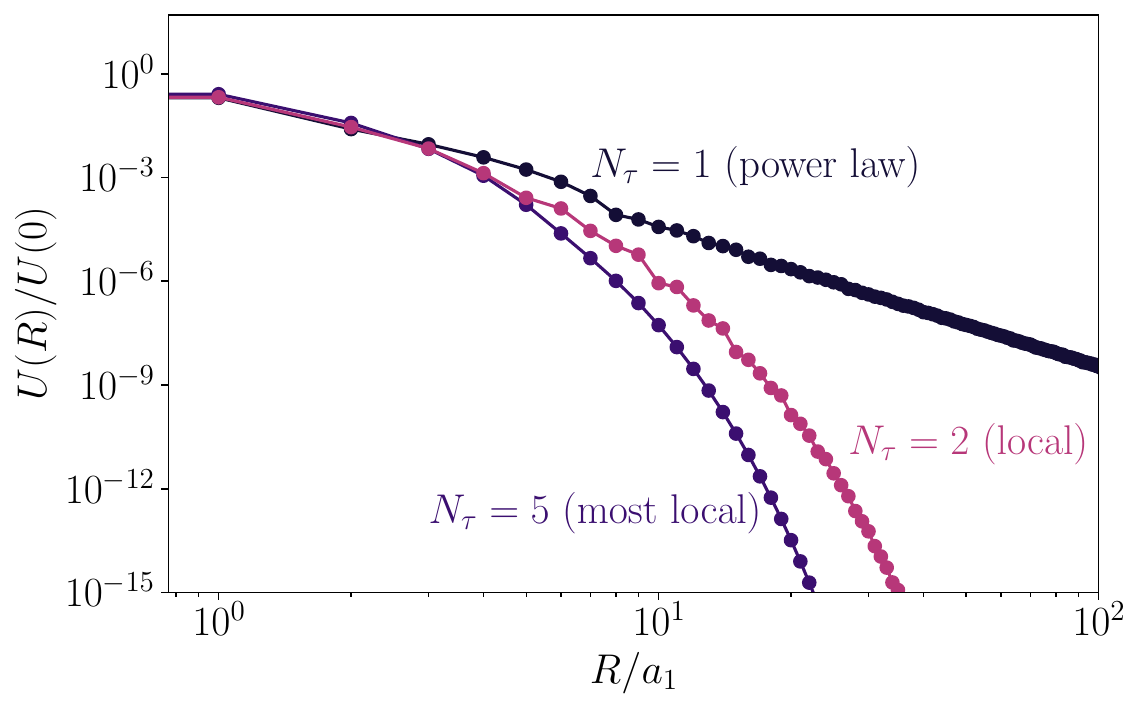}
  \caption{Real-space projected density-density interaction $U(R)$ along
  $R = n\,\mathbf{a}_1$ for the lattice spin-$S{=}2$ multifold
  ($|\mathcal{C}|=4$, five CNO modes), with the contact interaction of
  Appendix~\ref{app:coarse_grained_interaction}, normalized to the onsite
  value $U(0)$. The three curves label the three choices of truncation. The leading mode
  alone ($N_\tau=1$) gives a power-law tail, because its envelope zero
  leaves $\mathcal{S}_\bk$ singular and $\mathcal{S}_\bk^{-1/2}$
  non-analytic. The complete five-mode set ($N_\tau=5$) has constant
  $\mathcal{S}_\bk = 1$ and is the most localized. The pair $\{1,5\}$
  ($N_\tau=2$) is short-ranged but less localized than the complete set,
  since mode $5$ is finite where mode $1$ vanishes, keeping
  $\mathcal{S}_\bk$ nonzero but momentum dependent. The decay rate is set
  by the analyticity of $\mathcal{S}_\bk^{-1/2}$. A nonvanishing
  $\mathcal{S}_\bk$ gives exponential decay, a zero gives a power-law tail (for instance, when $N_\tau=1$).}
  \label{fig:range_mechanism}
\end{figure}

Retaining only the leading mode gives
$\mathcal{S}_\bk = \lambda_1|s_{1,\bk}|^2$. The leading envelope
$s_{1,\bk}$ has a topologically protected zero, so $\mathcal{S}_\bk$
has a singularity. At that momentum $\mathcal{S}_\bk^{-1/2}$ diverges, and by the argument above the projected interaction acquires a power-law tail. Any pair drawn from the first
four modes fails in the same way, because these four envelopes share a
zero at the same momentum, so the pair leaves $\mathcal{S}_\bk$ singular at that point. Adding orbitals does not help unless one of them is finite where the others vanish.

Mode $5$ is finite exactly where mode $1$ vanishes, so the pair has
\begin{equation}
\mathcal{S}_\bk^{\{1,5\}}
= \lambda_1|s_{1,\bk}|^2 + \lambda_5|s_{5,\bk}|^2,
\end{equation}
which is bounded away from zero across the whole Brillouin zone. The singularity is removed with two orbitals rather than five, and the
interaction is short-ranged. 
It is not, however, as local as the complete set. The reason is that $\mathcal{S}_\bk^{\{1,5\}}$ is nonzero
but not constant. 
Mode $5$ fills the gap in the neighborhood of the zero of mode $1$, but $\mathcal{S}_\bk^{\{1,5\}}$ retains a $\bk$-dependence that the complete set
does not have, so the decay is exponential but with a longer range.
The orbitals carry a residual real-space spread set by how sharply
$\mathcal{S}_\bk$ dips near the covered zero.

The three cases are described in Fig.~\ref{fig:range_mechanism}. The
leading mode alone is power-law because its envelope zero makes
$\mathcal{S}_\bk^{-1/2}$ non-analytic on the real Brillouin zone. The pair $\{1,5\}$ and the complete set are both exponential because their $\mathcal{S}_\bk$ never vanishes, with the complete set the more local of the two because its $\mathcal{S}_\bk$ is constant.

Similar phenomena occurs in chiral TBG.
The rank-1 truncation leaves
$\mathcal{S}_\bk$ singular because of the topological zero of the
leading envelope. Including the obstruction-covering CNO regularizes the
Gram matrix (Fig.~\ref{fig:gram_conditioning}) and restores
short-ranged projected interactions (Fig.~\ref{fig:range_of_interaction}).
Increasing the number of orbitals therefore improves locality not because the orbitals are very localized, but because the truncated orbitals provides a more faithful representation of the Bloch states.
The locality of the orbital function is therefore not sufficient.
One can pick ad hoc local states that are more localized than the CNO (such as a delta function centered at AA site), but such choices do not generally guarantee the smoothness of ${\rm det}[\mathcal{S}_\bk]$ in the Brillouin zone.

\section{Conclusions and Outlook}\label{secconclusions}

By diagonalizing the unit-cell reduced density matrix $L$, one obtains CNOs $\{|\tau_a\rangle\}$ with occupation numbers 
$\{\lambda_a\}$ that quantify the contribution of each mode to the charge density. This construction remains well-defined regardless of 
topological obstructions to exponentially localized Wannier functions.

There are three central results. First, the CNO decomposition is the
optimal rank-$N_\tau$ approximation to the density inside a unit cell,
with truncation error $\sum_{a>N_\tau}\lambda_a^2$. Second,
the leading envelope $s_{1,\bk}$ is a section of a line bundle of Chern number $\mathcal{C}$, so for a topological band it must vanish in the
Brillouin zone with total vorticity $2\pi\mathcal{C}$;  this forces more than one CNO for any Chern band and $S_\mathrm{cell}>0$. Third, the occupation spectrum sets a hierarchy in projected interactions.

The CNOs may enable more efficient many-body calculations by 
providing a natural truncation of the Hilbert space. For quantum 
embedding methods such as DMFT \cite{Georges1996}, CNOs offer 
a criterion for selecting impurity orbitals based on the charge density of the band. Applications to exact diagonalization 
and tensor network methods, where truncating to the leading $N_\tau$ 
CNOs would reduce computational cost, are promising directions for future work.

The CNO framework reframes the question of local orbital construction 
in topological systems. Rather than asking whether exponentially 
localized Wannier functions exist, we ask what the minimal local orbital 
structure consistent with the form factor structure of the band is. 
The answer, hidden in the spectrum and eigenmodes of $L$, bridges the gap between single-particle band topology and the real-space orbitals entering correlation physics.
The momentum-dependent spectral function induced by these principles should be measurable via the quantum twist microscope \cite{Inbar2023QTM, Birkbeck2025QTMPhonons, Xiao2026QTMFlatBands}. 

The spectrum outlines a new design principle. A band with exponentially decaying
CNO spectrum is dominated by a single interacting channel, with conventional
heavy-fermion phenomenology. 
On the other hand, a flat spectrum with multiple CNOs contributing to spectral weight has many competing channels. Whether tuning the spectrum this way can stabilize specific topological phases, fractional Chern insulators among them, is a promising direction. 
There are already some works relating charge density to the stability of FCIs in {\moire} systems \cite{shi2025effectsberrycurvatureideal, sarkar2026similarfractionalcherninsulators}.

\section*{Acknowledgments}
We gratefully acknowledge discussions with N.~Morales-Duran, E.~Khalaf, and P.~Ledwith. This material was supported by the National Science Foundation (NSF) under CAREER Award No.~DMR-2340394, the Materials Research Science and Engineering Centers
(MRSEC) program through Columbia University under the
Precision-Assembled Quantum Materials (PAQM) Grant No. DMR-2011738; Sloan Research Fellowship (FG-2025-24714), and the Army Research Office under Grant No.~W911NF-26-1-A283. The Flatiron Institute is a division of the Simons Foundation.

\appendix 

\begin{appendix}

\section{Periodic embedding in Continuum Models}
\label{app:periodic_continuum}
Continuum models expand the Bloch eigenstate in plane-wave states
$|\bk+\bG,l\rangle$, where $\bG$ runs over moir\'e reciprocal lattice
vectors up to a cutoff and $l$ labels internal degrees of freedom,
\begin{equation}
|\psi_{\bk,n}\rangle
=
\sum_{\bG,l}
z_{n,\bk,\bG,l}\,
|\bk+\bG,l\rangle,
\label{eq:app_psi_cont}
\end{equation}
with normalization $\sum_{\bG,l}|z_{n,\bk,\bG,l}|^2 = 1$.
Periodicity of the Bloch state,
$|\psi_{\bk+\bG',n}\rangle = |\psi_{\bk,n}\rangle$, requires
\begin{equation}
z_{n,\bk+\bG',\bG,l} = z_{n,\bk,\bG+\bG',l}.
\label{eq:app_z_shift_cont}
\end{equation}
The cell-periodic state in the physical gauge,
$|u^{\rm phys}_{\bk,n}\rangle = e^{-i\bk\cdot r}|\psi_{\bk,n}\rangle$ is written as
\begin{equation}
|u^{\rm phys}_{\bk,n}\rangle
=
\sum_{\bG,l}
z_{n,\bk,\bG,l}\,
|\bG,l\rangle.
\label{eq:app_uk_phys_cont}
\end{equation}
This wavefunction satisfies $|u^{\rm phys}_{\bk+\bG',n}\rangle = e^{-i\bG'\cdot r}
|u^{\rm phys}_{\bk,n}\rangle \neq |u^{\rm phys}_{\bk,n}\rangle$ and is therefore not
periodic.

The periodic embedding is defined by stripping the phase using the
unit-cell label $\bR$ rather than the full position $\br = \bR+\br_0$,
\begin{equation}
\langle\bR+\br_0|u_{\bk,n}\rangle
\equiv
e^{-i\bk\cdot\bR}\,
\langle\bR+\br_0|\psi_{\bk,n}\rangle.
\label{eq:app_uP_cont}
\end{equation}
Periodicity follows since $e^{-i\bG'\cdot\bR} = 1$ for any
reciprocal lattice vector $\bG'$ and Bravais lattice vector $\bR$:
\begin{align}
\langle\bR+\br_0|u_{\bk+\bG',n}\rangle
&=
e^{-i(\bk+\bG')\cdot\bR}
\langle\bR+\br_0|\psi_{\bk+\bG',n}\rangle \nonumber \\
&=
e^{-i\bk\cdot\bR}
\langle\bR+\br_0|\psi_{\bk,n}\rangle \nonumber \\
&=
\langle\bR+\br_0|u_{\bk,n}\rangle.
\end{align}
To express $|u_{\bk,n}\rangle$ in terms of the plane-wave
coefficients, substitute Eq.~\eqref{eq:app_psi_cont} into
Eq.~\eqref{eq:app_uP_cont} and sum over $\bR$,
\begin{equation}
\langle\br_0,l|u_{\bk,n}\rangle
=
\sum_\bG
z_{n,\bk,\bG,l}\,
e^{i(\bk+\bG)\cdot\br_0}
\equiv
e^{i\bk\cdot\br_0}\,
z_{n,\bk,\br_0,l},
\label{eq:app_uP_r}
\end{equation}
where the real-space coefficients are
\begin{equation}
z_{n,\bk,\br_0,l}
\equiv
\sum_\bG
e^{i\bG\cdot\br_0}\,
z_{n,\bk,\bG,l},
\label{eq:app_z_FT_cont}
\end{equation}
which are the discrete Fourier transform of $z_{n,\bk,\bG,l}$ from
$\bG$-space to real space. The real-space positions $\{\br_0\}$ form 
the discrete dual grid to the $\bG$-grid, satisfying
\begin{equation}
\frac{1}{N_G}
\sum_\bG
e^{i\bG\cdot(\br_0-\br_0')}
=
\delta_{\br_0,\br_0'},
\label{eq:app_dual_grid}
\end{equation}
which guarantees orthonormality of the states 
$|e_{\br_0,l}\rangle = N_k^{-1/2}\sum_\bR|\bR+\br_0,l\rangle$ 
and unitarity of the transform Eq.~\eqref{eq:app_z_FT_cont}.

The uc-RDM in Eq.~\eqref{eqrho_cell_BZ_average} is then computed
from the periodic-gauge amplitudes $e^{i\bk\cdot\br_0}
z_{n,\bk,\br_0,l}$,
\begin{equation}
L_{(\br_0,l),(\br_0',l')}
=
\frac{1}{N_k}
\sum_{\bk,n\in P}
e^{i\bk\cdot(\br_0-\br_0')}\,
z_{n,\bk,\br_0,l}\,
z^*_{n,\bk,\br_0',l'},
\label{eq:app_L_cont}
\end{equation}
which is an $N_G N_l \times N_G N_l$ Hermitian positive semidefinite
matrix. The $N_G$ real-space positions play the role of orbitals in 
direct analogy with the tight-binding case, and diagonalizing $L$ 
yields the CNOs and occupations as defined in the main text.

\section{Cell entropy and the quantum metric}
\label{app:entropy-cumulants}

The cell entropy $S_{\rm cell}$ is bounded below by the quantum distance
between Bloch states, averaged over the Brillouin zone. We derive this
bound here.

Write the uc-RDM as $L = \Pi P\Pi$, with $P$ the
band projector of Eq.~\eqref{eq:def_P_L} and $\Pi$ the cell projector
of Eq.~\eqref{eq:def_P_0}. On the reference-cell orbital space $L$
is an $n_{\rm orb}\times n_{\rm orb}$ matrix with eigenvalues
$\lambda_a\in[0,1]$, and the entropy is $S_{\rm cell}=\Tr f(L)$ with
\begin{equation}
f(x)=-x\ln x-(1-x)\ln(1-x).
\label{eq:Scell_app}
\end{equation}
The entropy obeys the elementary bound
\begin{equation}
f(x)\ge 4\ln 2\,\,x(1-x),
\end{equation}
saturated at $x=1/2$, where both sides equal $\ln 2$. Applied to each
eigenvalue,
\begin{equation}
S_{\rm cell}\ge 4\ln 2\,\big(\Tr L-\Tr L^2\big),
\end{equation}
so the entropy is controlled by the second charge cumulant
$\Tr L-\Tr L^2$~\cite{Klich2006,KlichLevitov2009,SongRachel2012,Kruchkov2025}.

This cumulant is a commutator norm. Using $L=\Pi P\Pi$
and $\Pi^2=\Pi$, $P^2=P$,
\begin{align}
\Tr L-\Tr L^2
&= \Tr(\Pi P) - \Tr(\Pi P\Pi P)\nonumber\\
&= \Tr\big[\Pi(1-P\Pi)P\big]\nonumber\\
&= \dfrac{1}{2} \big\|[\Pi,P]\big\|_{\rm HS}^2,
\end{align}
where $\|A\|_{\rm HS}=\sqrt{\Tr(A^\dagger A)}$ is the Hilbert-Schmidt norm. The cell entropy is large when the cell partition and the band projector do not commute.

In terms of cell-periodic states $|u_\bk\rangle$, this commutator norm is the Brillouin-zone average of
the quantum distance, giving
\begin{equation}
S_{\rm cell}\ge 4\ln 2\int_{\rm BZ}
\frac{d^dk}{(2\pi)^d}\frac{d^dk'}{(2\pi)^d}
\big(1-|\langle u_\bk|u_{\bk'}\rangle|^2\big).
\label{eq:C2dist}
\end{equation}
The bound ties the real-space entanglement of the cell to a global,
gauge-invariant measure of how far the states spread over the projective Hilbert space.

\section{Haldane model}\label{app:Haldane_model}
We outline the details of the Haldane model used in Sec.~\ref{sec1pRDM}. 
The model consists of two triangular sublattices $A$ and $B$. The Bravais 
lattice is spanned by the primitive vectors
\begin{equation}
\mathbf a_1= \Big(-\dfrac{\sqrt{3}}{2},\dfrac{3}{2}\Big),\quad
\mathbf a_2=\Big(\dfrac{\sqrt{3}}{2},\dfrac{3}{2}\Big),
\end{equation}
so that the corresponding reciprocal lattice vectors are
\begin{equation}
\mathbf b_1=\Big(-\dfrac{2\pi}{\sqrt{3}},\dfrac{2\pi}{3}\Big),\quad
\mathbf b_2=\Big(\dfrac{2\pi}{\sqrt{3}},\dfrac{2\pi}{3}\Big).
\end{equation}
In the $(A,B)$ sublattice basis, the Bloch Hamiltonian takes the 
$2\times 2$ form
\begin{equation}
H(\mathbf k)=
\begin{pmatrix}
g(\mathbf k, -\phi)+M & f(\mathbf k)\\[2pt]
 f^*(\mathbf k) & g(\mathbf k, \phi)-M
\end{pmatrix},
\end{equation}
where the nearest-neighbor and next-nearest-neighbor hoppings are
\begin{align}
f(\mathbf k) &= -t_1(1+e^{i\mathbf k\cdot\mathbf a_1}+e^{i\mathbf 
k\cdot\mathbf a_2}), \\
g(\mathbf k, \phi) &= -2t_2 \;[\cos(\mathbf k\!\cdot\!\mathbf 
a_1+\phi) \nonumber\\
&\phantom{=} + \cos(\mathbf k\!\cdot\!\mathbf a_2-\phi) \nonumber \\
&\phantom{=} + \cos(\mathbf k\!\cdot\!(-\mathbf a_1+\mathbf 
a_2)+\phi)].
\end{align}
Here $t_1$ and $t_2$ denote the hopping amplitudes; for simplicity 
we set $t_1=t_2=1$. The parameter $M$ corresponds to a staggered 
sublattice potential breaking inversion symmetry, and $\phi$ makes 
the hopping complex, producing a staggered magnetic flux breaking 
time-reversal symmetry without a net magnetic field.

The phase diagram is controlled by the competition between inversion 
breaking and time-reversal breaking. A nontrivial Chern insulating 
phase occurs when
\begin{equation}
|M| < 3\sqrt{3} t_2 |\sin\phi|,
\end{equation}
in which case the system realizes a quantum anomalous Hall state with 
Chern number $\mathcal{C}=\pm 1$. Outside this region the spectrum is 
topologically trivial ($\mathcal{C}=0$). In the main text 
(Fig.~\ref{fighaldane_zeros}), we focus on three representative points: 
$(M,\phi)=(0,\pi/2)$ and $(1, 2\pi/3)$ in the topological phase, and 
$(M,\phi)=(1,0)$ as a trivial insulator.

We use the trivial embedding where the Bloch Hamiltonian satisfies 
periodicity
\begin{equation}
H(\mathbf k+ \mathbf G) = H(\mathbf k), \quad \mathbf G = m\mathbf 
b_1+n\mathbf b_2,\quad m,n\in\mathbb Z.
\end{equation}
The location of the zero of $s_{1,\bk}$ reflects the underlying 
topological character of the band. When inversion symmetry is 
preserved ($M=0$), the zero is pinned at the $M$ point by symmetry. 
Breaking inversion symmetry ($M \neq 0$) displaces the zero away from
high-symmetry points, though its existence remains topologically
guaranteed as long as the system remains in the topological phase.

\section{Multi-fold Fermion model}\label{app:MultiFold_fermion}

Motivated by the multifold fermion \cite{wieder2016double, bradlyn2016beyond}, we consider the continuum Hamiltonian
\begin{equation}
    H_\bk = d_\bk \cdot \sigma, \quad 
    d_\bk = \left( k_x, k_y, \frac{k^2}{2M} - 1 \right),
\end{equation}
where $\sigma = ( \sigma^x, \sigma^y, \sigma^z )$ denotes the spin-$S$ representation of the $SU(2)$ Lie algebra, with dimension $2S+1$. Explicitly, the matrices satisfy
\begin{equation}
    [ \sigma^a, \sigma^b ] = i \epsilon^{abc} \sigma^c, 
    \quad \sum_a (\sigma^a)^2 = S(S+1).
\end{equation}
For a given $S$, the Hamiltonian $H_k$ thus acts in a $(2S+1)$ dimensional Hilbert space and possesses $2S+1$ eigenvalues,
\begin{equation}
    \lambda_{m}(\bk) = m \, |d_\bk|, \quad m = -S, -S+1, \dots, S,
\end{equation}
where $|d_\bk| = \sqrt{k_x^2 + k_y^2 + \left(k^2/2M+1\right)^2}$. The top band corresponds to $m=S$ and carries Chern number $\mathcal{C} = 2S$ \cite{Tan2024}.
The eigenstates of $H_\bk$ can be written as spin-coherent states oriented along $d_\bk$. Explicitly, the highest-wave-function of the top band ($m=S$) is
\begin{equation}
    | u_\bk \rangle = \frac{1}{(1+|z_\bk|^2)^S} 
    \sum_{m=0}^{2S} \sqrt{\binom{2S}{m}} \, z_\bk^m \, | S, S-m \rangle,
\end{equation}
where $|S, m\rangle$ are eigenstates of $\sigma^z$, and $z_\bk$ is the stereographic coordinate of $d_\bk$,
\begin{equation}
    z_\bk = \frac{d_x + i d_y}{|d_\bk| + d_z}.
\end{equation}
This parametrization is equivalent to a spin-$S$ generalization of the Bloch sphere construction.
The overlaps between eigenstates at different momenta define the form factors. Using the coherent state representation, one finds
\begin{equation}
    \langle u_\bk | u_{\bk'} \rangle 
    = \left( \frac{1+ z_\bk^* z_{\bk'}}{\sqrt{(1+|z_\bk|^2)(1+|z_{\bk'}|^2)}} \right)^{2S}
\end{equation}
from which the CNO spectrum can be computed.

\section{The truncation error equals the discarded eigenvalue weight}
\label{app:error_proof}

Here we give a self-contained derivation of the result
$\mathcal{E}_{\min}=\sum_{a>N_\tau}\lambda_a^2$
stated in Sec.~\ref{subsec:optimal_truncation}.
The argument requires only the spectral decomposition of $\Lambda$ and
the orthonormality of the eigenmodes $s_{a,n,\bk}$.

The overlap kernel and its rank-$N_\tau$ truncation are
\begin{align}
\Lambda_{n \bk,n'\bk'}
&=
\sum_{a\ge 1}\lambda_a\,s_{a,n,\bk}\,s^*_{a,n',\bk'},
\nonumber \\
\Lambda^{(N_\tau)}_{n \bk,n'\bk'}
&=
\sum_{a=1}^{N_\tau}\lambda_a\,s_{a,n,\bk}\,s^*_{a,n',\bk'},
\end{align}
with real eigenvalues $\lambda_1\ge\lambda_2\ge\cdots\ge 0$.
The eigenmodes satisfy
\begin{equation}
\sum_{n}\int_{\rm mBZ}dk\;
s_{a,n,\bk}\,s^*_{b,n,\bk}
=\delta_{ab}.
\label{eq:app_ortho}
\end{equation}
We write $\int_{n,\bk}\equiv\sum_n\int_{\rm mBZ}dk$ as a shorthand
for the combined sum-and-integral over the composite index $(n,k)$.

Because $\Lambda^{(N_\tau)}$ retains the first $N_\tau$ spectral terms,
\begin{equation}
\Lambda_{n \bk,n'\bk'}-\Lambda^{(N_\tau)}_{n \bk,n'\bk'}
=
\sum_{a>N_\tau}
\lambda_a\,s_{a,n,\bk}\,s^*_{a,n',\bk'}.
\label{eq:app_diff}
\end{equation}
Substituting into the error definition,
\begin{equation}
\mathcal{E}
=
\int_{n,\bk}\int_{n',\bk'}
\left|
\sum_{a>N_\tau}
\lambda_a\,s_{a,n,\bk}\,s^*_{a,n',\bk'}
\right|^2.
\label{eq:app_E1}
\end{equation}

Expanding with dummy indices $a$ and $b$,
\begin{equation}
\mathcal{E}
=
\int_{n,\bk}\int_{n',\bk'}
\sum_{a>N_\tau}\sum_{b>N_\tau}
\lambda_a\lambda_b\,
s_{a,n,\bk}\,s^*_{a,n',\bk'}\,
s^*_{b,n,\bk}\,s_{b,n',\bk'}.\nonumber
\end{equation}
The integrand factorises into a function of $(n,\bk)$ times a
function of $(n',\bk')$, so the double integral separates:
\begin{align}
\mathcal{E}
=
\sum_{a,b>N_\tau}
\lambda_a\lambda_b
\left(\int_{n,\bk} s_{a,n,\bk}\,s^*_{b,n,\bk}\right)
\left(\int_{n',\bk'} s^*_{a,n',\bk'}\,s_{b,n',\bk'}\right). \nonumber
\end{align}
Both factors in the parenthesis equal $\delta_{ab}$ by
orthonormality~\eqref{eq:app_ortho}.
We thus get
\begin{equation}
\mathcal{E}
=
\sum_{a>N_\tau}\lambda_a^2
\end{equation}
proving that the error is determined entirely by the discarded eigenvalues, so
retaining the largest $\lambda_a$ first minimizes $\mathcal{E}$ for
any fixed $N_\tau$.

\section{Projected Interactions}
\label{app:projectedInteractions}

For clarity we consider a single band per spin, dropping the band
index throughout.
Starting from a microscopic density-density interaction,
\begin{equation}
H_\mathrm{int}
=
\frac{1}{2}
\int d\br\,d\br'
\sum_{\sigma\sigma'}
n_{\br,\sigma}\,
V(\br-\br')\,
n_{\br',\sigma'},
\end{equation}
where $n_{\br,\sigma} = \psi^\dag_{r,\sigma} \psi^\phdag_{r,\sigma}$ is the density operator written in terms of field operators.
We expand the field operator in plane waves,
\begin{equation}
\psi^{\vphantom\dagger}_{\br,\sigma}
=
\int \frac{d\bq}{(2\pi)^2}\,
e^{i\bq\cdot\br}\,
\psi^{\vphantom\dagger}_{\bq,\sigma},
\end{equation}
to write the interaction as
\begin{equation}
H_\mathrm{int}
=
\frac{1}{2}
\int \frac{d\bq}{(2\pi)^2}\,
V(\bq)
\sum_{\sigma\sigma'}
\rho_{\bq,\sigma}\,\rho_{-\bq,\sigma'},
\end{equation}
where we have introduced
\begin{equation}
    \rho_{\bq,\sigma} =
\int \frac{d\bq'}{(2\pi)^2}\,
\psi^\dagger_{\bq'+\bq,\sigma}\,
\psi^{\vphantom\dagger}_{\bq',\sigma}.
\end{equation}
No assumption about a lattice has been made; as a result, all momenta are unrestricted.

A periodic lattice splits momentum $\bq$ into a crystal momentum $\bk\in\mathrm{BZ}$ and a reciprocal lattice vector $\bG$,
with $\bq = \bk+\bG$.
The plane-wave operators are related to Bloch band operators by
\begin{equation}
\psi^{\vphantom\dagger}_{\bk+\bG,\sigma}
=
\sum_n
z_{n,\bk,\bG}\,
c^{\vphantom\dagger}_{\bk,n,\sigma},
\label{eq:app_psi_bloch}
\end{equation}
which is the inverse of the plane-wave expansion
Eq.~\eqref{eq:app_psi_cont}.
Projecting to a single band $n=n$ and writing
$z_{\bk,\bG} \equiv z_{n,\bk,\bG}$,
\begin{equation}
\psi^{\vphantom\dagger}_{\bk+\bG,\sigma}
\;\to\;
\bar{\psi}^{\vphantom\dagger}_{\bk+\bG,\sigma}
=
z_{\bk,\bG}\,
c^{\vphantom\dagger}_{\bk,\sigma}.
\end{equation}
Substituting into the density operator and converting the momentum
integral to a discrete sum over $\bk\in\mathrm{BZ}$ and $\bG$,
\begin{align}
\bar{\rho}_{\bq,\sigma}
&=
\frac{1}{N_k}
\sum_{\bk,\bG}
\bar{\psi}^\dagger_{\bk+\bG,\sigma}\,
\bar{\psi}^{\vphantom\dagger}_{\bk+\bq+\bG,\sigma}
\nonumber \\
&=
\sum_{\bk}
\Lambda^{\rm phys}_{\bk,\bk+\bq}\,
c^\dagger_{\bk,\sigma}\,
c^{\vphantom\dagger}_{\bk+\bq,\sigma},
\end{align}
where the form factor is
\begin{equation}
\Lambda^{\rm phys}_{\bk,\bk+\bq}
=
\frac{1}{N_k}\sum_\bG
z^*_{\bk,\bG}\,
z_{\bk+\bq,\bG}
=
\frac{1}{N_k}\langle u_\bk|u_{\bk+\bq}\rangle,
\label{eq:app_Lambda_pw}
\end{equation}
and the last equality follows from the definition of the
physical-embedding cell-periodic state Eq.~\eqref{eq:app_uk_phys_cont}.
The projected interaction is
\begin{align}
\bar{H}_\mathrm{int}
&=
\frac{1}{2A}
\sum_{\bq\in {\rm all}}
V(\bq)
\sum_{\bk,\bk'\in\mathrm{BZ}}
\sum_{\sigma\sigma'}
\Lambda^{\rm phys}_{\bk,\bk+\bq}\,
\Lambda^{\rm phys}_{\bk',\bk'-\bq}\times \nonumber \\
&\hspace{3cm} c^\dagger_{\bk,\sigma}\,
c^{\vphantom\dagger}_{\bk+\bq,\sigma}\,
c^\dagger_{\bk',\sigma'}\,
c^{\vphantom\dagger}_{\bk'-\bq,\sigma'},
\label{eq:app_Hproj}
\end{align}
where $A$ is the total system area \cite{Kwan2025_moire_review}.

\section{All-to-all Density-Density Interactions}
\label{app:coarse_grained_interaction}
The general interaction framework in Sec.~\ref{secCNObasis} handles 
arbitrary microscopic interactions that include all momentum transfers $\bq$. 
Here we examine a limiting case where 
the interaction couples only unit-cell densities, restricting momentum transfers to reciprocal lattice vectors $\bG$. This yields a 
particularly simple limit where interaction scales are proportional to CNO eigenvalues.

Consider an interaction of the form
\begin{equation}
H_{\rm int} = \sum_{\bR,\bR'} V_{\bR-\bR'} \, \rho_\bR \rho_{\bR'},
\label{eq:Hint_coarse}
\end{equation}
where $\rho_\bR$ is 
the total electron density in unit cell $\bR$
\begin{equation}
    \rho_\bR = \sum_{r \in {\rm u.c.}} c^\dagger_{\bR+\br} c^\phdag_{\bR+\br} = \sum_{\br \in {\rm u.c.}} \rho_{\bR+\br} \label{eq:def_rho_R}
\end{equation}
This form should be viewed as a theoretical limiting case rather than 
a microscopic starting point. Physically, it corresponds to a screened 
Coulomb kernel that varies smoothly on the scale of the unit cell, 
with short-wavelength components exponentially suppressed.

Moving onto projected density operator, we use the Bloch wavefunctions for band $n$ to write it as
\begin{equation}
    \bar{\rho}_{\bR+\br} = \sum\limits_{\bk, \bk'} \psi_{\bk, n}(\bR+\br) \psi_{\bk', n}^*(\bR+\br) c^\dag_{\bk, n} c^\phdag_{\bk', n}.
\end{equation}
Using the periodic embedding defined in Eq.~\eqref{eq:app_uP_cont}, $\psi_{\bk, n}(\bR+\br) = e^{i \bk\cdot \bR} u_{\bk, n}(r)$, we get
\begin{equation}
    \bar{\rho}_{\bR+\br} = \dfrac{1}{N_k} \sum\limits_{\bk, \bk'} e^{ i (\bk - \bk')\cdot\bR} u_{\bk, n}(r) (u_{\bk',n}(r))^* c^\dag_{\bk, n} c^\phdag_{\bk', n}
\end{equation}
from which one can get the total density inside one unit cell using Eq.~\eqref{eq:def_rho_R} that results in
\begin{equation}
    \bar{\rho}_{\bR} = \dfrac{1}{N_k} \sum\limits_{\bk, \bk'} e^{ i (\bk - \bk')\cdot\bR} \langle u_{\bk, n} | u_{\bk',n} \rangle c^\dag_{\bk, n} c^\phdag_{\bk', n}.
\end{equation}
Substituting $\langle u^{\rm per}_{n, \bk} | u^{\rm per}_{n, \bk'} \rangle = \sum_a \lambda_a s_{a,\bk} s^*_{a,\bk'}$ gives
\begin{equation}
    \bar{\rho}_{\bR} = \sum\limits_a \lambda_a \; c^\dag_{\bR,a} c^\phdag_{\bR,a}
\end{equation}
where $c^\dag_{\bR,a}$ is the creation operator for the CNO $a$ in unit cell $\bR$.
With this decomposition, the projected interaction takes a remarkably simpler form
\begin{align}
    \bar{H}_{\rm int} = \sum_{\bR,\bR^\prime,a,b} V_{\bR-\bR^\prime}\, 
    \lambda_a \lambda_b\, 
    c^\dag_{\bR,a} c^\phdag_{\bR,a} 
    c^\dag_{\bR^\prime,b} c^\phdag_{\bR^\prime,b}.
\end{align}
The eigenvalues $\lambda_a$ thus set the interaction scale in 
each CNO. In the cases of interest there is a dominant mode 
$\lambda_D$ and a subdominant mode $\lambda_S$ associated with the 
band-inverted orbital, giving a natural hierarchy of interaction scales $U_f = V\lambda_D^2 \;>\; U_{cf} = V\lambda_D\lambda_S \;>\; 
    U_c = V\lambda_S^2$ where $U_f$, $U_c$, and $U_{cf}$ denote the intra-dominant, 
intra-subdominant, and cross-channel interactions respectively.
We note in passing that although this simplified form relies on the specific form of interaction, the hierarchy persists for realistic Coulomb interactions.

\section{Momentum-dependent Correlation Profile: Two-band model}
\label{app:momentum_selective_correlation}

We compute the single-particle spectral function for a flat Chern band with CNO-selective interactions in the Hubbard-I approximation. This strong-coupling limit treats each CNO site as an isolated atomic impurity and lets us see how an envelope zero affects the correlation gap. 
The approximation ignores quasiparticle renormalization, spatial correlations, and vertex corrections; its diagonal-in-CNO self-energy drops the inter-channel coupling.

Consider a flat Chern band with energy $\varepsilon_-$ and two CNOs $\{|\tau_1\rangle, |\tau_2\rangle\}$, with occupation numbers $\lambda_1 \gg \lambda_2$ satisfying $\lambda_1 + \lambda_2 = 1$. 
The noninteracting Hamiltonian projected into this two-CNO subspace is a $2\times 2$ matrix $h(\bk)$. The Bloch eigenstate $|u_{-,\bk}\rangle$ of this band decomposes in the CNOs as
\begin{equation}
|u_{-,\bk}\rangle = \sum_{a=1,2} \sqrt{\lambda_a} \, s^*_{a,\bk} |\tau_a\rangle,
\end{equation}
where the normalized envelope functions satisfy
\begin{equation}
\sum_a \lambda_a |s_{a,\bk}|^2 = 1.
\end{equation}
For a Chern band with $|\mathcal{C}| = 1$, topology enforces isolated zeros where $s_{1,\bk_0} = 0$ at certain $\bk_0$ in the Brillouin zone. The normalization constraint then implies $\lambda_2 |s_{2,\bk_0}|^2 = 1$, so spectral weight redistributes entirely to the subleading CNO.

We next consider onsite Hubbard interactions on both CNOs,
\begin{equation}
H_{\text{int}} = \sum_{a=1,2} U_a \sum_{\bR} n_{\bR,a,\uparrow} n_{\bR,a,\downarrow},
\end{equation}
where $n_{\bR,a,\sigma}$ is the density operator for CNO $a$ at unit cell $R$ with spin $\sigma$. 

Within a local approximation where self-energies remain diagonal in the CNOs, each channel develops a momentum-independent self-energy $\Sigma_a(\omega)$ determined by its local interaction $U_a$ and filling. The self-energy operator is $\Sigma(\omega) = \sum_{a=1,2} \Sigma_a(\omega) |\tau_a\rangle \langle \tau_a|$. Projecting onto the lower band gives
\begin{align}
\Sigma_{--}(\bk,\omega) &= \langle u_{-,\bk} | \Sigma(\omega) | u_{-,\bk} \rangle \nonumber \\
&= \sum_a \lambda_a |s_{a,\bk}|^2 \Sigma_a(\omega).
\label{eq:sigma_projection}
\end{align}
The momentum dependence comes entirely from the envelope structure, weighted by CNO occupations.

The Hubbard-I approximation treats each CNO site as an isolated impurity. 
The half-filled single-site Hubbard problem is particle-hole symmetric about its atomic level. Its atomic Green's function on CNO $a$ has poles at $\pm U_a/2$
\begin{equation}
G_{a,\text{atom}}(\omega) = \frac{1}{2} \left( \frac{1}{\omega - U_a/2} 
+ \frac{1}{\omega + U_a/2} \right)
\end{equation}
from which the Dyson equation gives the self-energy
\begin{equation}
\Sigma_a(\omega) = \omega - G^{-1}_{a,\text{atom}}(\omega) 
= \frac{U_a^2}{4\omega}.
\label{eq:hubbard_i}
\end{equation}
The pole at $\omega = 0$ generates upper and lower Hubbard bands at 
$\pm U_a/2$, separated by $U_a$. This is controlled in the 
strong-coupling limit $U_a \gg w$ where $w$ is the non-interacting
bandwidth.

Measuring frequencies relative to the chemical potential set at the 
flat band, the projected Green's function is
\begin{equation}
G_-(\bk,\omega) = \frac{1}{\omega - \Sigma_{--}(\bk,\omega)},
\end{equation}
which gives
\begin{equation}
\Sigma_{--}(\bk,\omega) = \sum_a \lambda_a |s_{a,\bk}|^2 
\frac{U_a^2}{4\omega}.
\end{equation}
from which one can calculate the spectral function $A_-(\bk,\omega) = -\text{Im}\, G_-(\bk,\omega + i\eta)/\pi$.
Now consider the rank-1 approximation where interactions act only on the leading CNO, setting $U_2 = 0$. The projected self-energy becomes
\begin{equation}
\Sigma_{--}(\bk,\omega) = \lambda_1 |s_{1,\bk}|^2 \Sigma_1(\omega).
\end{equation}
At the topological zero where $s_{1,\bk_0} = 0$, we get $\Sigma_{--}(\bk_0,\omega) = 0$ exactly for all $\omega$. The resulting spectral function at $\bk_0$ is a Lorentzian centered at $\omega = 0$, mimicking the Mott Semimetal phase \cite{Hofmann2022qmc, ledwith2024nonlocal, arx1_msi, arx2_msi, arx3_msi}.

The gapless crossing relies on the rank-1 structure of the projected
self-energy. Going beyond this level of approximation can generically restore the gap. The simplest is to include interactions on the subleading CNO ($U_2\neq0$), giving $\Sigma_{--}(\bk_0,\omega)=\Sigma_2(\omega)$
since $s_{1,\bk_0}=0$ and
$\lambda_2|s_{2,\bk_0}|^2=1$. The crossing is then replaced by a
correlation gap of order $U_2$.

More generally, any approximation that generates off-diagonal
self-energy components in the CNO basis or mixes the projected band with higher CNOs produces a finite
self-energy at $\bk_0$, even though the leading envelope vanishes.
Likewise, endowing momentum dependence to the self-energy
beyond the atomic Hubbard-I approximation can remove the exact decoupling responsible for the gapless crossing.
\end{appendix}

\bibliography{cno_refs}

\end{document}